\documentclass{applemlr}

\usepackage{amsmath}
\usepackage{enumerate}
\usepackage{algorithm}
\usepackage{algpseudocode}
\usepackage{amsfonts}
\usepackage{amsthm}
\usepackage{cleveref}
\usepackage{diagbox}
\usepackage{colortbl}
\usepackage{amssymb}
\usepackage{xspace}
\usepackage{wrapfig}
\usepackage{adjustbox}
\usepackage{tabularx}
\usepackage{booktabs}
\usepackage{mathtools}
\usepackage{tikz}
\usepackage{enumitem}
\usepackage{silence}
\usepackage{dsfont}
\usepackage[table]{xcolor}
\usepackage[dvipsnames]{xcolor}
\usepackage{multirow}
\usepackage{makecell}
\usepackage{xfakebold}

\usepackage{amsmath,amsfonts,bm}

\def\eqref#1{equation~\ref{#1}}

\def\1{\bm{1}}

\DeclareMathAlphabet{\mathsfit}{\encodingdefault}{\sfdefault}{m}{sl}
\SetMathAlphabet{\mathsfit}{bold}{\encodingdefault}{\sfdefault}{bx}{n}

\definecolor{textgray}{HTML}{6E6E73}
\usetikzlibrary{positioning, calc}
\usetikzlibrary{decorations.pathmorphing}

\makeatletter
\patchcmd{\wrong@fontshape}{\@gobbletwo}{}{}{}
\makeatother
\numberwithin{equation}{section}
\makeatletter
\AtBeginDocument{
  \urlstyle{sf}
  
}
\makeatother

\definecolor{light}{RGB}{125, 125, 125}
\crefname{tcb@cnt@pbox}{code}{code}
\Crefname{tcb@cnt@pbox}{Code}{Code}
\crefname{assumption}{assumption}{assumption}
\Crefname{assumption}{Assumption}{Assumptions}

\newtcolorbox[auto counter]{pbox}[2][]{
  colback=white,
  title=Code~\thetcbcounter: #2,
  #1,fonttitle=\sffamily,
  fontupper=\sffamily,
  arc=2pt,
  colframe=bgcolor,
  coltitle=fgcolor,
  colbacktitle=bgcolor,
  toptitle=0.25cm,
  bottomtitle=0.125cm
}

\makeatletter
\newcommand\applefootnote[1]{%
  \begingroup
  \renewcommand\thefootnote{}%
  \renewcommand\@makefntext[1]{\noindent##1}%
  \footnote{#1}%
  \addtocounter{footnote}{-1}%
  \endgroup
}
\makeatother

\definecolor{cverbbg}{gray}{0.90}

\makeatletter
\def\input@path{{TMLR26/}{TMLR26/sec/}{TMLR26/tables/}}
\makeatother

\graphicspath{{./}{TMLR26/}{TMLR26/figures/}}

\newcommand{\visatronic}{\texttt{Visatronic}\xspace}
\newcommand{\task}{VTTS\xspace}
\newcommand{\alignmetric}{\texttt{TimeSync}\xspace}

\definecolor{linkColor}{rgb}{0.18,0.39,0.62}
\newcommand{\contribblock}[2]{%
\vspace{0.2cm}
\begin{center}
\begingroup
\setlength{\fboxsep}{6pt}%
\fcolorbox{linkColor}{blue!8}{\parbox{0.94\linewidth}{\small\textbf{#1}~#2}}
\endgroup
\end{center}
\vspace{0.05cm}
}

\title{Mechanisms of Multimodal Synchronization: Insights from Decoder-Based Video-Text-to-Speech Synthesis}

\author[\dagger]{Akshita Gupta}
\author[*]{Tatiana Likhomanenko}
\author[*]{Karren Yang}
\author[*]{Richard He Bai}
\author[*]{Zakaria Aldeneh}
\author[*]{Navdeep Jaitly}

\affiliation[\dagger]{TU Darmstadt}
\affiliation[*]{Apple}

\abstract{
Unified decoder-only transformers have shown promise for multimodal 
generation, 
yet how sequence design affects the synchronization of modalities 
with heterogeneous sampling rates remains underexplored. 
We investigate 
this 
through video-text-to-speech (\task) synthesis---a 
controlled task requiring fine-grained temporal alignment between 
sparse text, video, and continuous speech. Using a unified decoder-only 
transformer, dubbed \visatronic{}, trained on VoxCeleb2, we 
 study: (i) how modalities contribute complementary 
information, (ii) how positional encoding strategies enable 
synchronization across heterogeneous rates, (iii) how modality ordering 
shapes the trade-off between in-domain performance and cross-domain 
transfer, (iv) how phoneme-level synchronization metrics provide 
diagnostic insight into per-phoneme timing errors. Our findings reveal that "global sequential indexing" (position IDs that count tokens in concatenation order) and "co-temporal ordered indexing" (identical IDs for temporally corresponding video and speech tokens) achieve comparable synchronization in-domain, whereas under a domain shift the two design axes interact: the combination of token ordering and position-ID scheme, rather than either alone, determines whether synchronization survives, with text-first co-temporal indexing transferring most robustly and requiring no explicit timestamp metadata. Both text 
and video contribute complementary signals: text ensures intelligibility 
while video provides temporal cues and emotional expressiveness. Modality ordering reveals a trade-off: video-first ordering achieves stronger in-domain performance, while among the timestamp-informed prefix variants text-first ordering generalizes more robustly to unseen domains. Training on diverse, in-the-wild data is associated with more 
transferable behavior, though isolating this factor requires targeted ablations.
To enable fine-grained analysis, we also introduce \alignmetric, 
a phoneme-level metric that reveals temporal misalignments overlooked 
by frame-level metrics.
These insights establish 
\task as a valuable testbed for understanding temporal synchronization 
in unified multimodal decoders. Generated speech results are attached in the supplementary.
}

\metadata[Code]{\url{{https://apple.github.io/visatronic-demo/}}}
\metadata[Correspondence]{\sffamily \url{akshita.gupta@tu-darmstadt.de}; \url{antares@apple.com}; \url{karren_yang@apple.com}; \url{richardbai@apple.com}; \url{zaldeneh@apple.com}; \url{njaitly@apple.com}}
\date{\sffamily\today}

\begin{document}

\maketitle

\applefootnote{\textcolor{textgray}{\sffamily Apple and the Apple logo are trademarks of Apple Inc., registered in the U.S. and other countries and regions.}}

\begin{figure*}[!ht]
    \centering
    \includegraphics[width=0.9\textwidth]{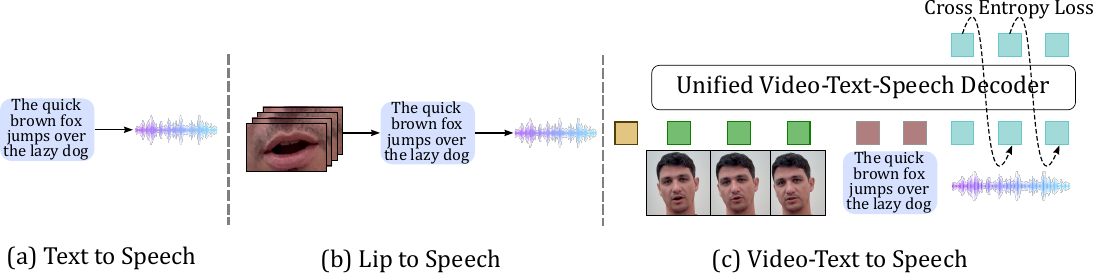}
    \caption{\textbf{\visatronic{} overview.} We study the video-text to speech (\task) task as a controlled playground to understand synchronization mechanisms in unified multimodal decoders. \visatronic{} is a unified decoder-only transformer that processes speaker tokens (yellow), video \textbf{v} (green), text \textbf{t} (red), and speech \textbf{s} (cyan) as discrete tokens in a shared sequence. We use this same color convention throughout the paper. The model is trained with cross-entropy loss $\mathcal{L}_{CE}$ to predict speech tokens, learning cross-modal interactions and temporal alignment across modalities with heterogeneous sampling rates (sparse text, 25fps video, 40Hz speech). By reducing confounding factors such as mixture-of-experts architectures and multi-stage training, VTTS enables systematic investigation of how unified decoders align temporal information. 
    }
    \label{fig:visatronic}
\end{figure*}

\section{Introduction}\label{sec:intro}

Unified decoder-only architectures have emerged as a powerful paradigm for 
multimodal generative tasks, demonstrating scalability across modalities~\citep{kondratyuk2024videopoet,parti,mm1,team2024gemini, achiam2023gpt, zhan2024anygpt, lu2024unified}.
Despite the empirical success of adapting pretrained large language models 
(LLMs) to multimodal inputs~\citep{xu2025qwen2, dai2023instructblip, yang2025qwen3, alayrac2022flamingo, su2023pandagpt, wu2024next}, 
how these models process and synchronize non-textual data across heterogeneous sampling 
rates remains poorly understood. While specialized models excel at generating 
individual modalities such as video~\citep{kondratyuk2024videopoet,yan2021videogpt} 
or speech~\citep{borsos2023audiolm}, unified LLMs may still struggle to map 
their linguistic strengths to fine-grained temporal dependencies in visual and 
acoustic domains~\citep{sakshi2024mmau,kondratyuk2024videopoet}. 
Deeper investigation is needed, not only into the performance of these
architectures, but also into the factors that impact decoder-only
multimodal synchronization performance across inputs and outputs. Throughout, we use \textit{mechanisms} to refer to the two factors we vary in isolation: position-ID assignment and modality ordering.


Understanding these mechanisms requires a relatively controlled setting that reduces 
synchronization from confounding architectural factors. Existing large-scale 
systems such as Qwen2.5-Omni~\citep{xu2025qwen25omnitechnicalreport} and 
AudioFlamingo~\citep{goel2025audio} operate at a scale where mixture-of-experts 
layers, multi-stage training pipelines, and massive dataset mixtures make it 
impossible to attribute synchronization behavior to specific design choices. 
For instance, key factors such as positional encoding design, modality
ordering, and handling heterogeneous frame rates can all affect
synchronization quality. Whether explicit timestamp-aware schemes are
necessary, or whether simpler implicit indexing mechanisms suffice,
cannot be answered by studying these highly complex systems alone.
A cleaner testbed is needed.

We use a minimal unified decoder, dubbed, \visatronic{} as an experimental platform deliberately avoiding mixture-of-experts routing and
multi-stage training, so that specific design choices---position-ID strategy and modality ordering can be varied in isolation. 
We also focus on a \textit{controlled ``playground'' task}: video-text to speech (VTTS) synthesis---generation of speech conditioned jointly on a video of a talking person and its corresponding text transcript, making synchronization quality directly measurable.
The ordering and
position-ID variants are the independent variables in our study, not
deployment-oriented engineering optimizations. VTTS is well suited for this
investigation for four reasons: (i) it is a representative generative task
with complementary multimodal input---text disambiguates homovisemes while
video provides temporal and expressive signals absent from text; (ii) it
involves heterogeneous sampling rates (sparse text, video frames, continuous
speech) from a synchronized source, making synchronization quality directly
measurable; (iii) successful generation requires fine-grained temporal
dependencies between visual articulatory cues and acoustic output; and (iv)
large-scale diverse datasets enable rigorous benchmarking. Our empirical investigation of
\visatronic{} on VTTS task yields four key findings.

\vspace{-0.2cm}
\paragraph{Decoder-only models can leverage complementary multimodal information.}
Systematic ablations show that, in a unified decoder-only architecture,
removing either text or video causes large degradation, confirming that
the model uses both modalities rather than collapsing to a single-source
shortcut. The contributions are asymmetric: text is more critical for
lexical content, while video provides temporal and expressive cues that
improve synchronization and can improve optimization efficiency under
specific ordering choices (Section~\ref{sec:exps}, Table~\ref{tab:modality_drop} and ~\ref{tab:voxceleb2_convergence}).

\vspace{-0.2cm}
\paragraph{Decoder-only models can synchronize heterogeneous sampling rates through position-ID design.}
As an alternative to explicit timestamp-aware schemes such as
TMRoPE~\citep{xu2025qwen25omnitechnicalreport}, we show that a unified
decoder-only model can bridge heterogeneous rates using standard positional
embeddings together with position-ID assignment. In our co-temporal ordered
indexing setup, video IDs are mapped to the speech-time index to provide a
shared temporal reference; in global sequential indexing, IDs follow
concatenation order without explicit cross-modal temporal correspondence
(Table~\ref{tab:voxceleb2_oldnewdata}). This shows that synchronization in
decoder-only VTTS is sensitive to position-ID design without
specialized alignment modules or explicit timestamp tokens in the model input.
We additionally show that modality ordering further affects synchronization,
with 
the optimal ordering depending on both evaluation metric and target domain: among the timestamp-informed prefix variants, text-first ordering generalizes more robustly across domains, while video-first ordering is stronger in-domain, indicating a trade-off between domain specialization and transferability (Section~\ref{sec:exps}, Tables~\ref{tab:lrs3_tab_obj} and ~\ref{tab:voxceleb2_oldnewdata}).

\vspace{-0.2cm}
\paragraph{Synchronization behavior transfers to out-of-domain data.}
\visatronic{}, trained exclusively on VoxCeleb2, generalizes zero-shot to 
LRS3--outperforming directly comparable models trained on LRS3 without domain-specific 
fine-tuning, lip detection, or architectural adaptation. 
This is consistent with the model learning transferable cross-modal structure rather than only dataset-specific shortcuts; we do not isolate this factor. 
Diverse training conditions (in-the-wild, noisy) may contribute to this transfer behavior. 
Furthermore, modality ordering affects generalization: among the timestamp-informed prefix variants, text-first ordering transfers more robustly to unseen videos than video-first ordering, consistent with lexical content being less sensitive to visual domain shift than appearance statistics
(Section~\ref{sec:sota_comp}, Table~\ref{tab:lrs3_tab_obj}).
\vspace{-0.2cm}
\paragraph{Evaluating decoder-only synchronization requires fine-grained metrics.}
To analyze synchronization behavior in a unified decoder-only model, we
introduce \alignmetric, a phoneme-level metric based on forced alignment
that measures absolute temporal offsets between corresponding phonemes in
generated and reference speech. This enables diagnosis of \textit{where}
and \textit{by how much} synchronization breaks down (e.g., ``the /s/
phoneme is 120ms early''). In contrast, widely used frame-level metrics
such as LSE-D/LSE-C~\citep{prajwal2020lip} and LMD~\citep{chen2019hierarchical}
produce aggregate scores and cannot localize errors to specific phonemes.
Thus, \alignmetric complements metrics such as WER by making decoder-only
temporal failure modes directly interpretable (Section~\ref{sec:time-sync},
Section~\ref{sec:exps} Figure~\ref{fig:metric_pdf} and
Table~\ref{tab:voxceleb2_oldnewdata}).




These findings come from one task family, two datasets, and models trained from scratch at $\sim$250M parameters; what we expect to carry beyond this setting is structural -- that position-ID assignment and modality ordering are not independent choices, and that the ordering that is best in-domain need not be the one that transfers.

We hope our findings provide actionable insights for designing unified decoder-only architectures for multimodal tasks requiring fine-grained temporal reasoning across heterogeneous modalities.
\section{\visatronic}
\label{sec: method}

\subsection{Video-Text To Speech (\task)}


Video-text-to-speech synthesis is formulated as follows. Given
(a) input video frames of a speaker
$\mathbf{x}^v \in \mathbb{R}^{T^v \times H \times W \times 3}$, where
$H$ and $W$ denote spatial video resolution (frame height and width, respectively) and $T^v$ is the total number of frames in the video;
and (b) text tokens $\{\mathbf{x}^t_i\}_{i=1}^{N}$, representing the transcript of speech in the video where
$\mathbf{x}^t_i \in \textit{Vocabulary}$ and $N$ is length of the tokenized transcript,
the goal is to generate
speech $\mathbf{x}^s \in \mathbb{R}^{T^s}$, where $T^s$ is length of speech signal, such that (i) spoken content
matches the text $\{\mathbf{x}^t_i\}_{1}^N$,
and
(ii) generated speech is temporally aligned with facial dynamics in video.

This task is challenging because it requires jointly modeling heterogeneous
modalities with different structures and sampling rates: sparse symbolic text,
video frames, and dense acoustic trajectories. These properties make VTTS a
suitable setting for testing synchronization mechanisms introduced in
Section~\ref{sec:intro}, especially the interaction between token ordering and
position-ID design. To isolate these effects, we adopt a simple unified
decoder-only modeling approach rather than specialized alignment modules.

\subsection{Discrete Input Representation Design}\label{sec:tokenization}

\begin{figure*}[h!]
    \centering
\includegraphics[width=0.8\textwidth]{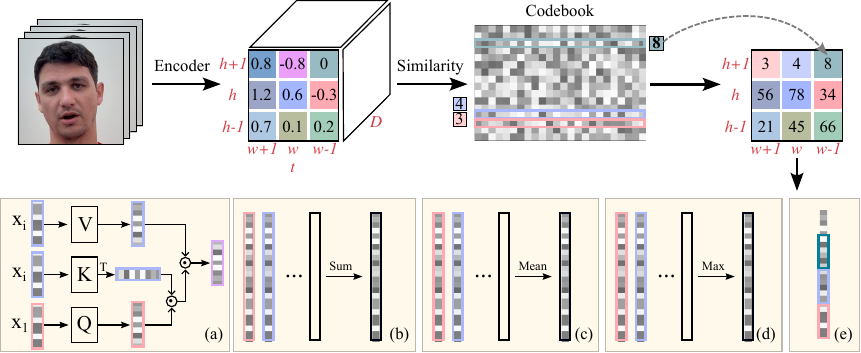}
    \caption{\textbf{Video representation.} Each video frame at time $t$ is encoded via a VQ-VAE~\citep{yan2021videogpt} into a downsampled spatial grid in $\mathbb{R}^{H'\times W'\times D}$. Each vector at location $(h, w)$ is quantized to a discrete token using the learned codebook $\mathbf{C}^v$ via $l_2$ similarity. These discrete tokens are embedded into $\mathbb{R}^{D'}$ and aggregated across the spatial grid to produce the final frame-level embedding input to the transformer. See Section~\ref{sec:tokenization} for details.}
    \label{fig:video_token}
\end{figure*}

\paragraph{Video Representation.}
To obtain a latent representation of the video input\footnote{In the rest of the paper, we denote tensors as $\mathbf{x}$ while $\mathbf{x}_{i, ...}$ denotes the $(i, ...)$-th component of tensor $\mathbf{x}$.} $\mathbf{x}^v$, we use a pretrained VQ-VAE model~\citep{yan2021videogpt}, trained on the Kinetics-600 dataset~\citep{carreira2018short}. Its reconstruction quality on our data is reported in Appendix ~\ref{app:video_recon}.
We choose VQ-VAE over discriminative video representations (e.g., CLIP~\citep{radford2021learning}, AV-HuBERT~\citep{shi2022learning}, or VideoMAE~\citep{tong2022videomae}) because VTTS requires preserving fine-grained spatial and temporal facial dynamics rather than semantic or 
category-level alignment. Discriminative encoders are optimized 
for global semantic matching or classification objectives and 
discard spatial locality in the process, whereas VQ-VAE retains 
a grid of discrete visual tokens per frame that encode fine-grained 
cues such as mouth shape, jaw motion, and facial muscle activations--all of which are critical for modeling phoneme timing and 
articulatory dynamics. Furthermore, VQ-VAE's reconstruction 
objective encourages preservation of visual detail at the frame 
level, making it better suited for a generation task where 
fine-grained temporal correspondence between visual and acoustic 
signals must be learned from scratch without pretrained 
audio-visual supervision.
Each frame $\mathbf{x}^v_t \in \mathbb{R}^{H \times W \times 3}$ is encoded to a spatially downsampled latent representation $\mathbf{y}^v_t \in \mathbb{R}^{H' \times W' \times D}$. Each spatial element in $\mathbf{y}^v_t$ is then quantized to the nearest entry in the VQ-VAE's learned codebook $\mathbf{C}^v=\{\mathbf{c}_1^v, \dots, \mathbf{c}_{K^v}^v\}$ using $\ell_2$ distance, resulting in a discrete token grid $\mathbf{v}_t \in [\mathbb{C}^v]^{H' \times W'}$. This quantization process is illustrated in Figure~\ref{fig:video_token}. 
The VQ-VAE compresses each $224 \times 224$ frame into a $16 \times 16$ grid using a codebook of size $K^v = 2048$ and embedding dimension $D = 3264$. These discrete tokens are then mapped to continuous embeddings $\mathbf{e}^v_{t,h,w} \in \mathbb{R}^{D'}$ via a learnable embedding layer $\mathbf{E}^v: \mathbb{C}^v \rightarrow \mathbb{R}^{D'}$, resulting in an embedding map $\mathbf{e}_t^v \in \mathbb{R}^{16 \times 16 \times D'}$. We explore multiple spatial aggregation functions over this grid (Section~\ref{sec:exps}, Table~\ref{tab:video_agg}; the functions are defined in Appendix ~\ref{app:video_rep}) and select summation based on the ablation results:
$\mathbf{z}_t^v = \sum_{h=1}^{16} \sum_{w=1}^{16} \mathbf{e}^v_{t,h,w}$.
This final vector $\mathbf{z}_t^v$ serves as a compact, information-rich embedding of the frame that preserves spatial and speaker-specific features while aligning to the decoder’s input dimension.

\vspace{-0.2cm}
\paragraph{Text Representation.}

For text processing, we employ a character-level tokenizer that maps the input text $\{\mathbf{x}^t_i\}_{1}^N$ to a sequence of discrete tokens $\mathbf{t}_j\in\mathbb{C}^t=\{1,2, \dots, K^t\}$ with $|\mathbb{C}^t| = K^t$, followed by a learnable embedding layer $\mathbf{E}^t(\cdot): \mathbb{C}^t \to \mathbb{R}^{D'}$. 
Character-level tokenization reduces vocabulary size $K^t$ and improves generalization by capturing fine-grained linguistic features.

\vspace{-0.2cm}
\paragraph{Speaker Representation.}

For multi-speaker modeling, we extract speaker representations using a pre-trained dvector model~\citep{variani2014deep} that produces 512-dimensional embeddings. These speaker embeddings are projected through a learnable linear layer to match the model dimension $D'$, and are used to
preserve speaker characteristics in generated speech.

\vspace{-0.2cm}
\paragraph{Speech Representation.}
\begin{figure*}[t!]
    \centering
\includegraphics[width=0.95\textwidth]{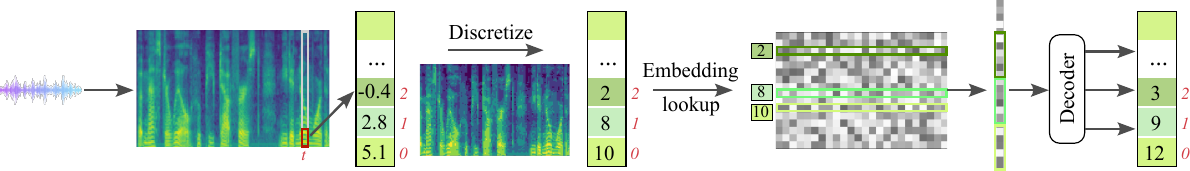}
    \caption{\textbf{Speech representation.} We follow the speech discretization process from dMel by~\cite{bai2024dmel}: each continuous mel-filterbank at time $t$ extracted from the raw audio is mapped into a discrete values using a codebook of evenly spaced values. 
    Afterwards, each discretized log mel-filterbank at time $t$ is mapped through a learnable embedding layer, all representations for log mel-filterbanks at time $t$ are stacked together and resulting vector is projected by a learnable linear layer to the model dimension~$D'$.
    All discretized log mel-filterbanks at time $t$ are predicted in parallel and independently.
    }
    \label{fig:speech_token}
\end{figure*}

We use dMel~\citep{bai2024dmel}, 
a discretization approach for speech processing; see Figure~\ref{fig:speech_token} for an overview. 
Given an input speech signal $\mathbf{x}^s$, we first compute continuous log mel-filterbanks $\mathbf{y}_{t}^s\in\mathbb{R}^F$ for a frame at time $t$, where $F$ is the number of log mel-filterbanks.
Then, we map every log mel-filterbank $\mathbf{y}_{t,f}^s\in\mathbb{R}$ to a discrete value $\mathbf{s}_{t, f}\in\mathbb{C}^s=\{1,2,\dots2^{K^s}\}$ using a codebook $\mathbf{C}^s=\{\mathbf{c}^s_1, \mathbf{c}^s_2,\dots, \mathbf{c}^s_{2^{K^s}}\}$: $\mathbf{c}^s_i\in\mathbb{R}$ are evenly spaced values in the range $[m, M]$, where $m$ and $M$ are the minimum and maximum values of mel-filterbanks computed across the dataset. To discretize, we take the closest codebook value, i.e, $\mathbf{s}_{t,f}=\text{argmin}_{i\in\mathbb{C}^s}|\mathbf{y}_{t,f}^s - \mathbf{c}^s_i|$. 
After each speech frame is discretized, every discrete value is mapped via a learnable embedding layer $\mathbf{E}^s(\cdot): \mathbb{C}^s \to \mathbb{R}^{d'}$ to a representation $\mathbf{e}^s_{t,f}$. The representation for the whole frame is given by $\mathbf{e}_t^s\in\mathbb{R}^{F\times d'}$, where $d'$ is the intermediate dimension. 
Subsequently, we stack these embeddings and project the resulting vector to a final embedding $\mathbf{z}^s_t\in\mathbb{R}^{D'}$ via a learnable linear layer $\mathbf{L}^s(\cdot): \mathbb{R}^{Fd'} \to \mathbb{R}^{D'}$: $\mathbf{z}_t^s = \mathbf{L}^s([\mathbf{e}^s_{t,1}, \mathbf{e}^s_{t,2}, \dots, \mathbf{e}^s_{t,F}])$.
This training-free discretization enables effective processing of speech signals in our framework.
Following~\cite{bai2024dmel} we use $K^s=4$ bits with $|\mathbb{C}^s|=16$, $F=80$ log mel-filterbank channels and $d'=24$. 

\vspace{-0.2cm}
\paragraph{Speech Inversion.}
To reconstruct the speech signal $\mathbf{x}^s$ from the speech discrete values $\mathbf{s}_{t, f}$ predicted by the multimodal transformer decoder (Section \ref{sec:multimodal_transformer}), we follow~\cite{bai2024dmel}: first, we transform the indices back to the log mel-filterbanks via the codebook $\mathbf{C}^s$: $\mathbf{\hat{y}}^s_{t,f}=\mathbf{c}^s_{\mathbf{s}_{t,f}}$. 
Subsequently, we apply a vocoder~\citep{yamamoto2020parallel} to transform reconstructed log mel-filterbanks $\mathbf{\hat{y}}^s_{t,f}$ back into the time domain signal~$\mathbf{x}^s$. 
The vocoder is trained independently and is not part of the \visatronic model.

\subsection{Unified Multimodal Video-Text-Speech Transformer Decoder}\label{sec:multimodal_transformer}

We use a unified multimodal decoder-only transformer architecture for processing multiple modalities~-- video, text and speech -- in order to generate speech given video and text inputs, see Figure~\ref{fig:visatronic}. The architecture consists of a single transformer decoder that processes \textbf{discrete} multimodal input representations from Section~\ref{sec:tokenization}. Unlike traditional approaches that use one modality as input, or separate encoder(s) for multimodal input, our unified architecture enables cross-modal interactions through self-attention layers while maintaining temporal coherence.


While we leverage general-purpose pretrained models for modality-specific preprocessing (VQ-VAE for video encoding~\citep{yan2021videogpt}, dvector for speaker embedding~\citep{variani2014deep}, and vocoder for speech reconstruction~\citep{yamamoto2020parallel}), these components remain frozen and serve purely as feature extractors. Critically, our approach does not require task-specific audio-visual encoders (e.g., AV-HuBERT~\citep{shi2022learning}), emotion recognition networks, or explicit fusion modules as in recent video-to-speech methods~\citep{choi2025v2sflow}. Instead, all trainable parameters reside in a single unified decoder that jointly learns cross-modal alignment and speech generation through self-attention, without additional task-specific alignment modules.

The model is trained end-to-end using cross-entropy loss to predict in parallel all channels of the next discrete token in sequence, allowing it to learn intrinsic relationships across modalities that are crucial for tasks requiring multimodal understanding. 
During inference, the model can generate tokens autoregressively while maintaining coherence across all modalities.

\begin{figure}[t!]
    \centering\includegraphics[width=0.6\columnwidth]{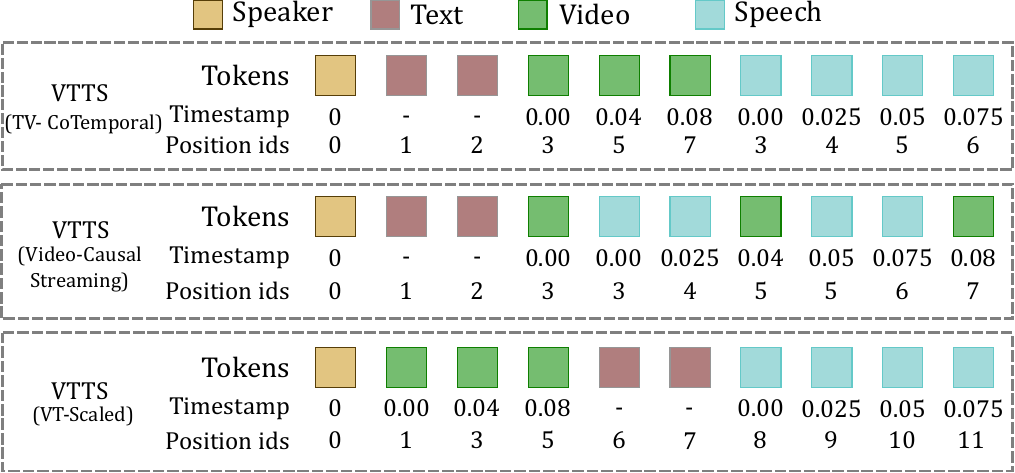}
    \caption{\textbf{Input sequence for \visatronic.} We encode all modalities into a discrete token space (see Figures~\ref{fig:video_token} and~\ref{fig:speech_token}), which is directly consumed by the decoder-only transformer. Each modality’s discrete representation is indicated by a colored square. Each row illustrates a different token-order strategy for combining multimodal information: (top) TV-CoTemporal, where text precedes video, then speech, with true video-speech position-ID overlap for co-temporal tokens; (middle) Video-Causal-Streaming, where text appears first while speech and video are interleaved in temporal order such that speech generation at time $t$ attends to the full text and only past video tokens at $t' \le t$; (bottom) VT-Scaled, where video precedes text, then speech, with timestamp-scaled video IDs but no direct video-speech ID overlap. Global sequential counterparts (TV-Global, VT-Global) are shown in Figure~\ref{fig:token_mix_global}. Reading a row here against its global-ID counterpart in Figure~\ref{fig:token_mix_global} isolates the position-ID computation at fixed token order; reading the top two rows (TV-CoTemporal and Video-Causal-Streaming) against each other isolates token order at fixed IDs.}
    \label{fig:token_mix}
\end{figure}

\vspace{-0.2cm}
\paragraph{Integration of Multimodal Sequences.}
For effective processing of multiple modalities with different temporal resolutions, we implement various input mixing strategies, see Figure~\ref{fig:token_mix}. Systematically exploring these strategies allows us to understand how token ordering and temporal alignment affect cross-modal learning in unified decoders. The fundamental challenge lies in handling different sampling rates and temporal ordering: speech inputs from dMel are sampled at 25ms intervals $(0.00s, 0.025s, 0.05s, \dots)$, whereas 25fps video inputs are sampled at 40ms intervals $(0.00s, 0.04s, 0.08s, \dots)$, and text tokens appear sparsely in the sequence.
We explore three strategies for combining multimodal sequences, differing 
in both token ordering and position ID assignment. These two dimensions—ordering and position indexing—are separately controllable design axes that we vary systematically to study both their individual and interacting effects (see Figures~\ref{fig:token_mix} and~\ref{fig:token_mix_global}).

\paragraph{Sequence parameterization.} 
Every sequence in our framework is specified by three per-token quantities, shown as the three rows of Figure~\ref{fig:token_mix}: (i) the token order -- where each modality's tokens sit in the sequence, subject to the constraint that speech never precedes the tokens it is conditioned on; (ii) the timestamp -- the true time of the token (video every 40ms, speech every 25ms; text and speaker tokens carry no timestamp); and (iii) the position ID -- the index passed to RoPE, computed either from the timestamp or from sequence order alone. Choosing these three quantities produces each of the five variants we evaluate: the three canonical orderings and the two global-ID ablations. TV-CoTemporal computes video and speech IDs from timestamps on a shared axis, so tokens at the same instant share an ID; VT-Scaled scales only the video IDs by timestamp and numbers text and speech sequentially after them, so video and speech never share an ID;
TV-Global and VT-Global keep the same two prefix orders but ignore timestamps and number tokens by sequence order; Video-Causal-Streaming orders video and speech tokens by their timestamps. Table~\ref{tab:variants} in Appendix~\ref{app:variants} summarizes the token order, position-ID computation, and whether video and speech share IDs for each variant; together with each variant's causal constraint and intended setting, while intermediate schemes expressible in the same terms are discussed in Appendix~\ref{app:limitations}.

The parameterization also covers configurations between the two extremes. Video-Causal-Streaming sits between full-prefix conditioning and full temporal interleaving, and its timing-error distribution behaves accordingly (Figure~\ref{fig:metric_pdf}).

Unlike timestamp-aware position encodings such as TMRoPE~\citep{xu2025qwen25omnitechnicalreport}, we add no timestamp tokens to the model input and use no alignment module. The model runs standard RoPE; timestamps enter only through position-ID assignment and, in the streaming variant, through token order. Comparing the co-temporal scheme against the global-ID variants therefore measures what this implicit temporal information contributes (Section~\ref{sec:exps}).

\vspace{-0.2cm}
\begin{itemize}[leftmargin=3mm] 
\item \textbf{Prefix strategies with timestamp-informed IDs (TV-CoTemporal, VT-Scaled).}
These variants use full-prefix conditioning (all non-speech tokens appear
before speech) but differ in ordering and position-ID semantics.
\textbf{TV-CoTemporal} (text $\rightarrow$ video $\rightarrow$ speech) uses true
co-temporal overlap: video and speech tokens that correspond in time are
assigned the same position IDs, creating a shared temporal axis without
explicit timestamp tokens. \textbf{VT-Scaled} (video $\rightarrow$ text
$\rightarrow$ speech) uses timestamp-scaled video IDs; text IDs occupy
contiguous positions inside the conditioning prefix, 
and speech
IDs start sequentially after the full prefix. The ordering
defines the autoregressive history available at each decoding step and
influences cross-modal dependency learning with implications for both
in-domain performance and cross-domain generalization (Section~\ref{sec:exps}).

\item \textbf{Global Sequential Indexing (TV-Global, VT-Global).}
For each prefix ordering above, we evaluate a global-indexing ablation.
Here, all tokens receive strictly increasing position IDs by concatenation
order. This is the simplest available scheme: the modality blocks are concatenated and position IDs simply count tokens in order. This scheme does not encode cross-modal temporal correspondence in
position IDs, allowing direct comparison against TV-CoTemporal and
VT-Scaled.

\item \textbf{Streaming (Video-Causal-Streaming).}
Text tokens appear first, followed by video and speech tokens 
interleaved in their original temporal order, such that the speech 
token at time $t$ attends only to past video tokens at $t' \le t$, i.e., the current and past frames. 
This enforces strict causal constraints between video and speech--the model cannot attend to future video frames when generating speech--reducing attention overhead while preserving temporal progression of both modalities. Position IDs follow the temporal order, 
reflecting the natural arrival order of tokens in a streaming setting. Current systems handle differing sampling rates by interleaving tokens in time rather than by repeating them; Qwen2.5-Omni~\citep{xu2025qwen25omnitechnicalreport} follows the same pattern, cited here as a design reference rather than as a baseline, since it is a large pretrained model trained on massive data mixtures and not comparable to our from-scratch setup. Text remains a prefix in all variants because it has no sampling rate of its own.

\end{itemize}
\vspace{-0.4cm}


\paragraph{Which orderings are admissible.} The model is a causal decoder trained to predict speech tokens conditioned on the other modalities, so every conditioning token must precede the speech tokens; placing speech before video and text is not usable for conditional speech generation, since the speech tokens would have no access to the video or text they are generated from. The reversal that is compatible with causal decoding is interleaving speech with video in temporal order, which is Video-Causal-Streaming: speech at time $t$ attends to video at $t' \le t$, so video still precedes each speech token. Reversal within the conditioning prefix is what we study directly---TV and VT are the two orders of text and video, evaluated under both ID schemes. The only excluded reversal is placing speech before its conditioning.

\vspace{-0.1cm}
\paragraph{Positional Encoding.}
With sequences longer than typical TTS and heterogeneous sampling 
rates across modalities, positional encoding design becomes critical. 
Prior work has consistently shown that relative positional embeddings 
outperform absolute embeddings for long 
sequences~\citep{touvron2023llama,bai2024dmel}. We apply 
RoPE~\citep{su2024roformer} across the entire sequence for its 
extrapolation properties and efficiency. Together with the global-ID ablations, the variants above let us vary position-ID assignment at fixed token order and token order at fixed position-ID scheme. \\

\vspace{-0.3cm}
\paragraph{Initialization.}
Placing all modalities' inputs into one sequence for the decoder, we found that having different submodules to map each modality to the shared space leads to inconsistency of the embeddings across modalities (e.g. they have very different norm magnitudes). 
Thus, proper initialization of these submodules is essential.
We identified that a proper scale for the initial weights distribution by bringing all inputs' final embeddings to the same sphere is sufficient for stable and fast training convergence. \\
\vspace{-0.2cm}
\paragraph{Model Training.}
Our unified decoder model is trained to predict speech discrete representations while being conditioned on all modalities during inference. 
During training we compute the cross-entropy loss $\mathcal{L}_{CE}$ only on the speech discrete representations, omitting the loss on others. 
All $F$ discrete log mel-filterbanks at each timestamp $t$ are predicted independently and in parallel. To ensure robust training, we follow dMel training observations and apply random span masking with probability $p$ to video, text and speech representations, forcing the model to leverage cross-modal information rather than relying solely on one modality. 
Speech masked regions are excluded from the loss computation. 
During inference, the model autoregressively generates speech discrete representations while being conditioned on speaker information, video and text. 

\contribblock{Design Highlights:}{
(i) single unified decoder that jointly processes speaker, video, text, and speech discrete tokens without task-specific fusion modules; 
(ii) controlled sequence-design decomposition into token ordering (TV-CoTemporal, VT-Scaled, Video-Causal-Streaming) and position-ID assignment (co-temporal overlap, timestamp-scaled prefix IDs, global sequential IDs), enabling clear ablations of synchronization mechanisms; 
(iii) robust training with random span masking and speech-only cross-entropy supervision, which discourages single-modality shortcuts and is intended to promote cross-modal dependency learning; 
(iv) embedding-scale initialization that aligns modality projection magnitudes and is used to improve optimization stability and convergence speed.}

\section{\alignmetric}\label{sec:time-sync}
To evaluate how well the generated speech aligns temporally with the ground truth, we define a phoneme-level alignment metric, \alignmetric. It is computed as:
\begin{equation}
\alignmetric = \frac{1}{N}\sum_{\phi^{GT}} \left| t^{model}_{f(\phi^{GT})} - t^{GT}_{\phi^{GT}} \right| 
\end{equation}
where $N$ is the total number of phonemes $\phi^{GT}$ in the ground truth transcriptions obtained from the original audio samples. 
$t^{GT}_{\phi^{GT}} = (start_{\phi^{GT}} + end_{\phi^{GT}}) / 2$ denotes the segment's center (in seconds) for each ground truth phoneme. 
$f(\phi^{GT})$ denotes the corresponding aligned phoneme in the generated audio, and 
$t^{model}_{f(\phi^{GT})} = (start_{f(\phi^{GT})} + end_{f(\phi^{GT})}) / 2$ is the center of that segment. 
Here, $start$ and $end$ indicate the phoneme's start and end timestamps in either the generated or ground truth audio. 
This metric captures the temporal deviation between aligned phoneme centers and helps quantify how well the model preserves synchronization. Because each phoneme carries its own center time, utterance length never enters the unit of measurement and no master window is needed; measuring per phoneme is what allows the metric to localize errors, whereas frame-level metrics return a single aggregate score.

\begin{figure}[t!]
\centering
\begin{minipage}{0.48\textwidth}
    \centering
    \includegraphics[width=0.95\linewidth]{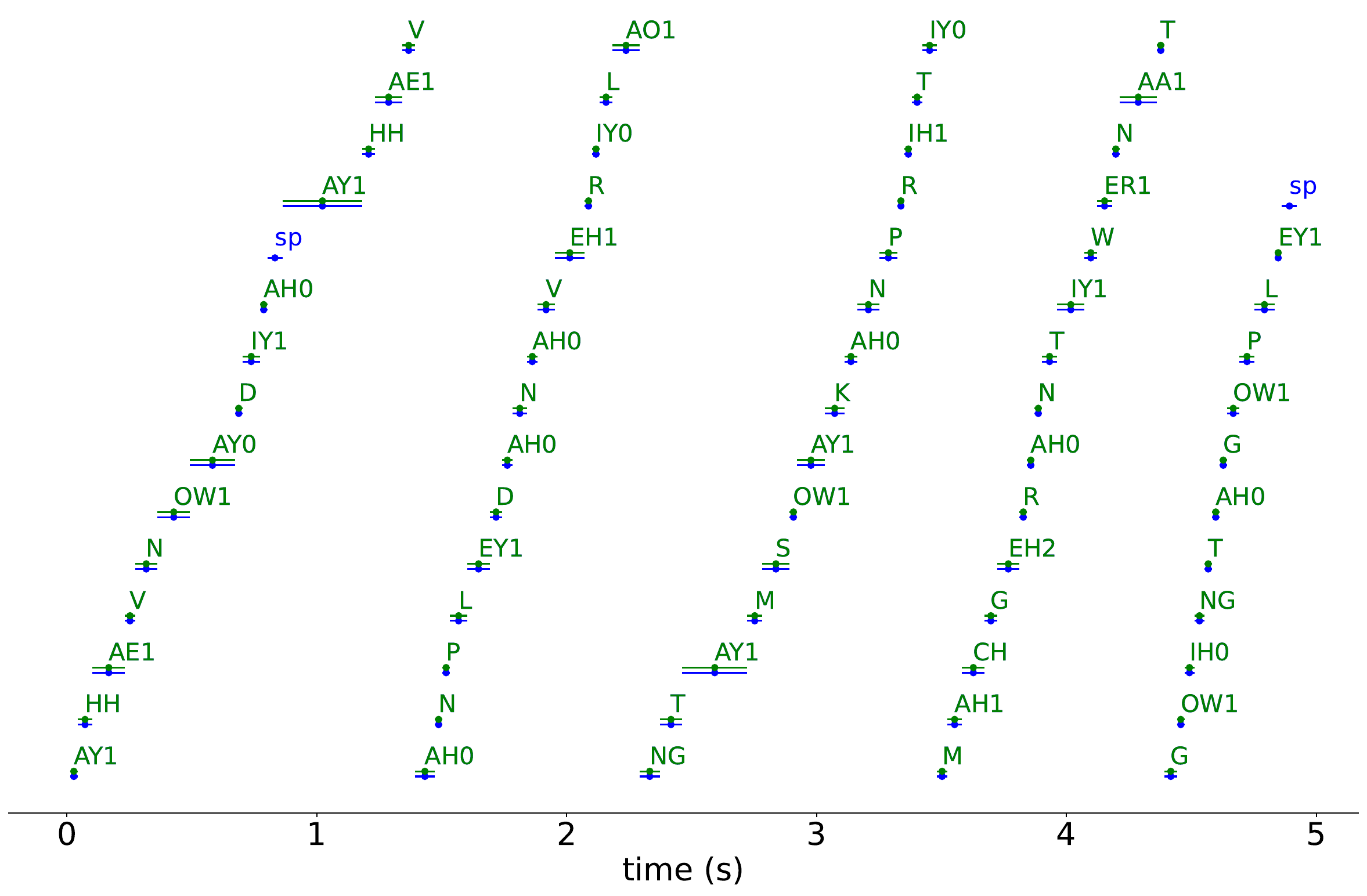}
\end{minipage}\hfill
\begin{minipage}{0.48\textwidth}
    \centering
    \includegraphics[width=0.95\linewidth]{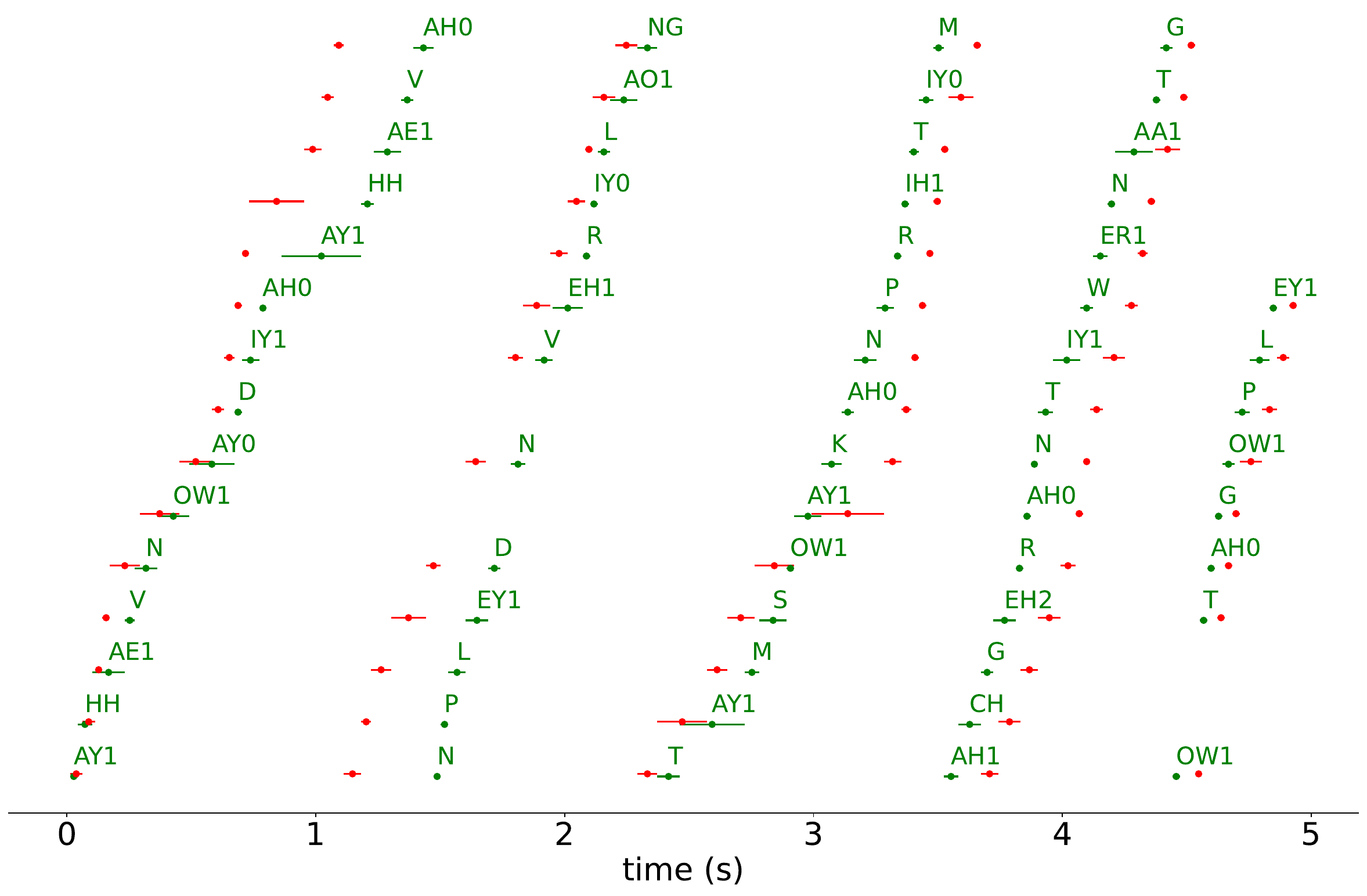}
\end{minipage}
\caption{\textbf{\alignmetric.} Visualization of phoneme-level alignment used for computing the \alignmetric. Left: alignment in ground truth audio before (blue) and after (green) removing silence (``sp'') segments. Right: aligned phoneme positions between ground truth (green) and generated (red) audio, where \alignmetric is computed as the absolute difference between segment centers (measured in seconds).}
\label{fig:alignmetric_combined}
\end{figure}

In practice, \alignmetric is computed by taking ground truth transcription and doing force alignment of its phoneme sequence to audio (either generated or original audio) 
via forced alignment using an HMM model (we use \url{https://github.com/richardbaihe/a3t} from \cite{pmlr-v162-bai22d}); we use the HTK toolkit~\citep{young2002htk} as our implementation, though the metric itself is toolkit-agnostic.
This procedure gives us phoneme location in time and phoneme duration for each audio. 
Afterward, we exclude silence (``sp'') and its duration from each alignment (Figure~\ref{fig:alignmetric_combined}, left), as our current metric focuses on spoken-phoneme timing rather than pause timing. Retaining silence durations would make a single pause rendered longer or shorter than the reference shift every phoneme after it, so that one error would be recounted for each following phoneme and would mask the phoneme timing itself. Pause behavior is instead measured by the duration and spectrogram results, where VTTS matches the ground truth's 393 frames while TTS overruns to 445 (Figure~\ref{fig:qual_mel_metric}, left; Figures~\ref{fig:qual_mel_suppl_v2} and~\ref{fig:qual_mel_suppl_164}).
Because every word has several possible phoneme sequences, we use Levenshtein distance to align phoneme sequences obtained for generated and original audio: we consider phonemes aligned if they are equal or related by substitution. Such pairs are kept in the timing computation rather than discarded: the aligner may return different but corresponding phonemes for the two audios even when the spoken word is correct. Only phonemes with no counterpart at all -- insertions and deletions -- are left out of the average. Then, we compute the average absolute time difference between centers of aligned phoneme segments in generated and original audio (Figure~\ref{fig:alignmetric_combined}, right), averaged over all phonemes in the test set, so frequent phonemes dominate by construction; a per-type average, weighting rare phonemes equally, is an alternative view of the same alignments.

LSE-D and LSE-C operate on lip-sync confidence at the frame level and return one aggregate score per utterance, so they cannot say which phoneme is misaligned or by how much. MCD-DTW-SL additionally allows time warping, which can absorb the very misalignment being measured (Appendix~\ref{app:subjective_evals}). We report \alignmetric{} because the question here is where synchronization breaks down, not whether a clip is broadly in sync. Sensitivity of the metric to its inputs, and its agreement with human synchronization judgments, are analyzed in Section~\ref{sec:timesync_anl} (\alignmetric Analysis).

\contribblock{\alignmetric Enables:}{(i) phoneme-level diagnosis of where synchronization breaks down; (ii) absolute temporal offset measurement without warping artifacts; (iii) fine-grained assessment of temporally salient misalignments, whereas frame-level aggregate metrics (e.g., LSE-D/LSE-C) do not provide phoneme-level localization.}
\section{Experiments}
\label{sec:exps}

We systematically evaluate \visatronic{} across four axes: modality 
contributions, positional encoding and ordering strategies, 
cross-domain generalization, and phoneme-level synchronization 
diagnosis. Our evaluation combines objective metrics 
(WER, \alignmetric), subjective human assessment (MOS), and ablation 
studies to understand not just \textit{what} the model achieves but 
\textit{how} performance varies with these design choices. Comparisons 
against prior work serve to confirm that cross-modal information flow 
works as intended, rather than to claim superiority over specialized 
systems. We train on VoxCeleb2 and evaluate zero-shot transfer to LRS3.

\vspace{-0.2cm}
\paragraph{Datasets.}
1) \textbf{LRS3}~\citep{afouras2018lrs3} is audio-visual dataset in English, compiled from TED and TEDx video presentations.
2) \textbf{VoxCeleb2}~\citep{chung18b_interspeech} is a large-scale audio-visual dataset primarily designed for speaker recognition task but applicable to various audio-visual processing domains.
Firstly, we develop a pipeline for pseudo-labeling (PL) the speech using Demucs \citep{defossez2020real} for speech enhancement, Whisper-large v2~\citep{radford2023robust} for automatic transcription, and proper data filtering as data are multilingual and without any text annotation. 
The initial version of labeled data, PL.v1, was obtained by keeping English-only detected samples. 
Later, we improved upon it by additional filtering of inconsistent too long or too short transcriptions, resulting in a cleaner PL.v2 version.
PL.v2 is treated as the ground truth for VoxCeleb2 in WER and alignment analyses, and its noise is quantified rather than assumed: GT WER of 2.6 \textpm 0.1\% (Table~\ref{tab:voxceleb2_oldnewdata}) is the Whisper transcription of the original audio scored against PL.v2, a direct bound on the transcript noise entering both. The effect of label quality is also measured: Table~\ref{tab:voxceleb2_oldnewdata} reports PL.v1 and PL.v2 rows, so the improvement from 4.0\% to 2.6\% GT WER and the corresponding improvement in model results are visible as a before/after. The conclusions do not depend on it: all compared variants share the same PL.v2 transcripts, so label noise cannot account for differences between variants, and the ordering trade-off, the video-conditioning benefit and the \alignmetric{} ranking all hold on LRS3, whose transcripts are human-annotated (Tables~\ref{tab:lrs3_mos},~\ref{tab:lrs3_tab_obj},~\ref{tab:lrs3_tab_obj_3to45}).

\vspace{-0.2cm}
\paragraph{Objective Evaluation Metrics.}
To evaluate generated speech quality, we use two objective metrics. 
\textbf{Word Error Rate (WER)} measures intelligibility by comparing 
Whisper-large v2 transcriptions of generated speech against ground 
truth transcripts, capturing how well the model preserves linguistic 
content. \textbf{\alignmetric} (Section~\ref{sec:time-sync}) measures 
phoneme-level temporal offsets through forced alignment of phoneme 
sequences to audio using an HMM model from HTK~\citep{young2002htk}, 
providing finer-grained synchronization diagnosis than established 
frame-level metrics such as LSE-D and LSE-C~\citep{prajwal2020lip}, 
which produce only aggregate scores and cannot identify which specific 
phonemes are misaligned or by how much. Together, WER and \alignmetric 
capture complementary aspects of generation quality: intelligibility 
and temporal synchronization respectively. We use this framing as a 
diagnostic interpretation rather than claiming calibrated perceptual 
correlation against frame-level metrics in this work.

\vspace{-0.2cm}
\paragraph{Subjective Evaluation Metrics.}
We randomly selected 50 samples each from VoxCeleb2 and LRS3 test data for human evaluation to assess the naturalness, intelligibility, 
synchronization and emotional expressiveness of the generated speech following~\cite{yemini2024lipvoicer}. Using mean opinion score (MOS) with 95\% confidence intervals, human evaluators rated the speech naturalness, intelligibility, synchronization and emotional expressiveness on a scale of 1 to 5, where 1 represents the worst and 5 the best quality. These are absolute ratings rather than A/B comparisons; the one comparative task is the emotion-similarity rating, in which two samples are played side by side. Details on the full protocol as well as implementation and training details are provided in Appendix~\ref{app:subjective_evals} and ~\ref{app:training_details}.

\begin{table}[t]
\caption{\textbf{Human evaluation on LRS3.} Mean opinion scores (MOS) (1-5) with 95\% confidence intervals for our VoxCeleb2-trained models evaluated on LRS3. VTTS (TV-CoTemporal) achieves better performance in Intelligibility (3.62) and Synchronization (3.12), while both methods perform equally well in Naturalness (3.01). GT is an upper bound, while GT (discrete) shows degradation due to speech quantization and vocoding artifacts. These results demonstrate our model maintains good perceptual quality even on out-of-distribution data. For baseline methods, we use publicly available model checkpoints from their respective repositories to generate speech samples, which we then evaluate using our human evaluation protocol. ``-'' indicates unavailable metrics under the same standardized setup; this applies to dubbing baselines where only naturalness scores were available.
}
\label{tab:lrs3_mos}
\centering
\small
\resizebox{0.7\columnwidth}{!}{
\begin{tabular}{@{}lcccccc@{}}\toprule
    \textbf{Method} & \textbf{Intelligibility}  ($\uparrow$) & \textbf{Naturalness}  ($\uparrow$) & \textbf{Synchronization}  ($\uparrow$) \\\midrule
    GT & 4.79 \textpm 0.05 & 4.79 \textpm 0.05 & 4.73 \textpm 0.06 \\ 
    GT (discrete) & 4.32 \textpm 0.11 & 3.80 \textpm 0.11 & 4.59 \textpm 0.07 \\
    \midrule
    \multicolumn{4}{@{}l}{\textbf{\textit{Lip to Speech Synthesis}}} \\
    Lip2Speech~\citep{kim2023lip} & 2.21 \textpm 0.08  & 2.20 \textpm 0.09  & 2.69 \textpm 0.08 \\
    SVTS~\citep{mira2022svts} & 2.17  \textpm 0.08  & 2.15 \textpm 0.09 & \underline{2.71 \textpm 0.09} \\
    VCA-GAN~\citep{kim2021lip} &  2.19 \textpm 0.08  & 2.20 \textpm 0.09 & \underline{2.71 \textpm 0.08} \\ \midrule
    \multicolumn{4}{@{}l}{\textbf{\textit{Dubbing Methods}}} \\
    HPMDubbing~\citep{cong2023learning} & -  & \underline{2.37 \textpm 0.14} & - \\
    StyleDubber~\citep{cong2024styledubber} & -  & 1.79 \textpm 0.12 & - \\ \midrule
    \multicolumn{4}{@{}l}{\textbf{\textit{VTTS}}} \\
    TV-CoTemporal & \textbf{3.62 \textpm 0.20} & \textbf{3.01 \textpm 0.22} & \textbf{3.12 \textpm 0.27 } \\
    VT-Scaled & \underline{3.30 \textpm 0.21} & \textbf{3.01 \textpm 0.17} & 2.35 \textpm 0.22 \\
\bottomrule
\end{tabular}
}
\end{table}

\begin{table}[t!]
\caption{\textbf{Human evaluation of emotional expressiveness on VoxCeleb2.} Mean Opinion Scores (1-5) with 95\% confidence intervals for emotional consistency between video and speech (\textit{video-speech emotions}), perceived emotion quality in speech alone (\textit{speech emotions}), and specific emotion categories (\textit{Angry, Happy, Fearful}). VTTS (TV-CoTemporal) achieves the highest score in video-speech emotion alignment (3.79), while VTTS (VT-Scaled) performs best in overall speech emotion quality (3.39) and in all individual emotion categories. These results demonstrate the advantage of visual conditioning for expressive speech generation. Ground truth (GT) serves as the reference, and GT (discrete) provides the upper bound achievable with our quantization-based approach. Per-category labels were collected only for GT, TTS, and VTTS (VT-Scaled) due annotation-budget constraints.}
\label{tab:voxceleb_mos_emotion}
\centering
\resizebox{0.8\columnwidth}{!}{
\begin{tabular}{@{}lcccccc@{}}\toprule
    \textbf{Method} & \textbf{video-speech emotions} ($\uparrow$) & \textbf{speech emotions} ($\uparrow$) & \textbf{Angry} ($\uparrow$) & \textbf{Happy} ($\uparrow$) & \textbf{Fearful} ($\uparrow$) \\\midrule
    GT & 4.62 \textpm 0.07 & 4.92 \textpm 0.04 & 4.36 & 4.43 & 4.20 \\
    GT (discrete) & 4.41 \textpm 0.10 & 4.37 \textpm 0.12 & -- & -- & -- \\ \midrule
    TTS (dmel ~\cite{bai2024dmel}) & 3.57 \textpm 0.14 & 3.20 \textpm 0.15 & \underline{3.46} & \underline{3.58} & \underline{3.17} \\ \midrule
    \multicolumn{6}{@{}l}{\textbf{\textit{VTTS}}} \\
    Video-Causal-Streaming & 3.66 \textpm 0.16 & \underline{3.36 \textpm 0.15} & -- & -- & -- \\
    TV-CoTemporal & \textbf{3.79 \textpm 0.15} & 3.31 \textpm 0.17 & -- & -- & -- \\
    VT-Scaled & \underline{3.74 \textpm 0.12} & \textbf{3.39 \textpm 0.15} & \textbf{3.82} & \textbf{3.76} & \textbf{3.73} \\
\bottomrule
\vspace{-0.4cm}
\end{tabular}
}
\end{table}


\subsection{Human Evaluation Results}
Human evaluation is presented in Table~\ref{tab:lrs3_mos},
where we compare our VTTS models against baselines from both lip-to-speech synthesis (Lip2Speech~\citep{kim2023lip}, SVTS~\citep{mira2022svts}, VCA-GAN~\citep{kim2021lip}) and dubbing literature (HPMDubbing~\citep{cong2023learning}, StyleDubber~\citep{cong2024styledubber}). Participants rated generated speech based on intelligibility, naturalness, and synchronization using 5-point mean opinion scores (MOS) with 95\% confidence intervals. Entries marked ``-'' indicate metrics that were unavailable for those systems under a standardized protocol.
Our VTTS (TV-CoTemporal) model achieves the highest scores among all generation systems: 3.62 in intelligibility and 3.12 in synchronization, and ties with VTTS (VT-Scaled) on naturalness (3.01). The synchronization gap between TV-CoTemporal (3.12) and VT-Scaled (2.35) is consistent with our ordering analysis: text-first conditioning yields stronger cross-domain synchronization robustness in human perception. These results are notably closer to the ground truth (GT) and GT (discrete) upper bounds, highlighting both the expressiveness and temporal coherence of our approach. 

\vspace{-0.2cm}
\paragraph{Emotional Expressiveness Evaluation.}
Beyond intelligibility, naturalness, and synchronization, we evaluate whether 
video conditioning enables emotionally expressive speech that aligns with 
facial expressions. We select a subset of 50 samples with strong non-neutral 
emotions (using emotion2vec~\citep{ma2024emotion2vec} model with >99\% confidence 
across 6 emotion classes) and conduct additional human evaluation. Raters assess two dimensions: (1) \textit{video-speech emotions:} how well 
the emotional tone in speech matches the facial expressions in the video, and 
(2) \textit{speech emotions:} the perceived emotional quality and expressiveness 
of the speech itself. Table~\ref{tab:voxceleb_mos_emotion} presents results 
across these dimensions and specific emotion categories (Angry, Happy, Fearful) for \textit{speech emotions}. VTTS (TV-CoTemporal) achieves the highest video-speech emotion alignment score among generation systems (3.79/5.0), notably outperforming TTS (3.57/5.0) which lacks visual 
context. For overall speech emotion quality, VTTS (VT-Scaled) achieves 
3.39/5.0 compared to TTS's 3.20/5.0, while VTTS (Video-Causal-Streaming) remains competitive (3.36/5.0), indicating that streaming constraints retain substantial emotional information. Notably, VTTS (VT-Scaled) demonstrates 
strong performance across all individually scored emotion categories (Angry: 3.82, 
Happy: 3.76, Fearful: 3.73), approaching the ground truth upper bound. This split between variants is consistent with our other measured trends: TV-CoTemporal is strongest on synchronization/alignment-focused metrics (LRS3 sync MOS in Table~\ref{tab:lrs3_mos} and video-speech emotion alignment in Table~\ref{tab:voxceleb_mos_emotion}), whereas VT-Scaled is strongest on in-domain speech quality/content metrics (VoxCeleb2 WER in Table~\ref{tab:voxceleb2_oldnewdata} and intelligibility/naturalness MOS in Table~\ref{tab:voxceleb2_mos}). We report per-category ratings only for GT, TTS, and the variant strongest on speech-emotion quality (VT-Scaled) to keep annotation load tractable. The confidence intervals of the VTTS variants overlap on both emotion dimensions, so we do not read a ranking among them. These results demonstrate that visual conditioning not only improves temporal 
synchronization but also enables the model to generate emotionally expressive 
speech that naturally aligns with speakers' facial expressions, a critical 
capability for applications like dubbing and character animation where emotional consistency is essential.

\begin{table*}[t!]
\centering
\caption{\textbf{Cross-dataset generalization and validation against prior work on full-set LRS3.}
Primary comparisons are reported on the full
LRS3 test set for fairness. VoiceCraft-Dub* is shown for reference only and is not directly
comparable because it adapts a large pretrained TTS backbone with additional AV-specific
fine-tuning.}
\label{tab:lrs3_tab_obj}
\resizebox{0.64\columnwidth}{!}{
\begin{tabular}{lcc}
\toprule
\textbf{Method} & \textbf{Training Dataset} & \textbf{WER (\%)} $\downarrow$ \\
\midrule
V2SFlow \citep{choi2025v2sflow} & LRS3 & 28.5 \\
LipVoicer \citep{yemini2024lipvoicer} & LRS3 & 21.4 \\
Lip2Speech \citep{kim2023lip} & LRS3 & 57.4 \\
SVTS \citep{mira2022svts} & LRS3 & 82.4 \\
VCA-GAN \citep{kim2021lip} & LRS3 & 90.6 \\
DiffV2S \citep{choi2023diffv2s} & LRS3 & 39.2 \\
VoiceCraft-Dub* (pretrained) \citep{sung2025voicecraft} & LRS3 fine-tuned & 1.38 \\
\midrule\midrule
TTS & VoxCeleb2 & \underline{27.8} \\
\midrule
\multicolumn{3}{@{}l}{\textbf{\textit{VTTS}}} \\
VT-Scaled & VoxCeleb2 & 59.6 \\
TV-Global & VoxCeleb2 & 45.1 \\
VT-Global & VoxCeleb2 & 38.5 \\
TV-CoTemporal & VoxCeleb2 & \textbf{17.9} \\
\midrule
VT-Scaled & LRS3 & 28.9 \\
TV-CoTemporal & LRS3 & \textbf{17.7} \\
\bottomrule
\end{tabular}
}
\end{table*}

\subsection{State-of-the-Art Comparison}
\label{sec:sota_comp}

\begin{table*}[t!]
\caption{\textbf{WER and \alignmetric on VoxCeleb2.} We report three 
WERs: (i) GT WER from original audio, (ii) GT (discrete) WER from 
reconstructed audio using GT speech tokens (lower bound), and (iii) 
WER from generated audio by a model. All values are computed using 
Whisper-large v2. The first two rows correspond to PL.v1 transcripts; 
the remaining rows use PL.v2. \alignmetric is reported for the PL.v2 section. Among
non-global prefix variants, VTTS (VT-Scaled) 
achieves 12.2\% WER, while global-indexing ablations reach 12.1--12.7\% WER. 
\alignmetric further shows improved phoneme-level 
synchronization with video-conditioned models. Rows labeled ``global'' compare 
Global Sequential Indexing as an ablation against 
their non-global counterparts (TV-CoTemporal and VT-Scaled).}
\label{tab:voxceleb2_oldnewdata}
\vspace{0.1cm}
\centering
\setlength{\tabcolsep}{6pt}
\resizebox{0.9\textwidth}{!}{%
\renewcommand{\arraystretch}{1.12}
\begin{tabular}{@{}lccccc@{}}\toprule
    \textbf{Method} & \textbf{Input Modality} & 
    \textbf{GT WER} ($\downarrow$) & 
    \textbf{GT (discrete) WER} ($\downarrow$) & 
    \textbf{WER} ($\downarrow$) & 
    \textbf{\alignmetric} (s) ($\downarrow$) \\ \midrule
    TTS (dmel~\cite{bai2024dmel}) & Text & 
        \multirow{3}{*}{4.0 \textpm{0.1}} & 
        \multirow{3}{*}{10.5 \textpm{0.1}} & 
        \underline{19.0}{\tiny{\textcolor{teal}{$\mathord+8.5$}}} & - \\
        \cmidrule(lr){1-2}\cmidrule(lr){5-6}
    \multicolumn{2}{@{}l}{\textbf{\textit{VTTS}}} & & & & \\
    VT-Scaled & Video-Text & 
         & 
         & 
        \textbf{17.2}{\tiny{\textcolor{teal}{$\mathord+6.7$}}} & - \\ 
    \midrule\midrule
    TTS & Text & 
        \multirow{4}{*}{2.6 \textpm{0.1}} & 
        \multirow{4}{*}{10.1 \textpm{0.2}} & 
        14.7{\tiny{\textcolor{teal}{$\mathord+4.6$}}} & 0.62 \textpm 0.98 \\
        \cmidrule(lr){1-2}\cmidrule(lr){5-6}
    \multicolumn{2}{@{}l}{\textbf{\textit{VTTS}}} & & & & \\
    Video-Causal-Streaming & Text-Video & 
         & 
         & 
        14.5{\tiny{\textcolor{teal}{$\mathord+4.4$}}} & 0.49 \textpm 0.63 \\
    LTTS (LT-Scaled) & Lip-Text & 
         & 
         & 
        14.0{\tiny{\textcolor{teal}{$\mathord+3.9$}}} & \underline{0.46 \textpm 0.61} \\ 
    \midrule
    \multicolumn{2}{@{}l}{\textbf{\textit{VTTS (Global IDs)}}} & 
        \multirow{3}{*}{2.6 \textpm{0.1}} & 
        \multirow{3}{*}{10.1 \textpm{0.2}} & 
        & \\
    TV-Global & Text-Video & 
        & 
        & 
        12.7{\tiny{\textcolor{teal}{$\mathord+2.6$}}} & 
        0.48 \textpm 0.69 \\
    VT-Global & Video-Text & 
         & 
         & 
        \textbf{12.1}{\tiny{\textcolor{teal}{$\mathord+2.0$}}} & 
        \underline{0.46 \textpm 0.61} \\ 
    \midrule
    \multicolumn{2}{@{}l}{\textbf{\textit{VTTS (Non-global IDs)}}} & 
        \multirow{3}{*}{2.6 \textpm{0.1}} & 
        \multirow{3}{*}{10.1 \textpm{0.2}} & 
        & \\
    TV-CoTemporal & Text-Video & 
        & 
        & 
        14.1{\tiny{\textcolor{teal}{$\mathord+4.0$}}} & \textbf{0.44 \textpm 0.65} \\
    VT-Scaled & Video-Text & 
         & 
         & 
        \underline{12.2}{\tiny{\textcolor{teal}{$\mathord+2.1$}}} & 0.47 \textpm 0.63 \\ 
\bottomrule
\end{tabular}
}
\end{table*}

To validate that our architectural choices properly leverage multimodal information, we 
compare against prior work. These comparisons serve to confirm that cross-modal information 
flow works as intended, rather than to claim superiority over specialized systems.
Comparisons in this paper fall into three kinds
: ablation comparisons inside our own framework, where every model shares the discretization, vocoder, architecture and recipe; directly comparable prior work evaluated on the same full LRS3 test set; and reference-only systems that differ in pretraining, representation or fine-tuning.

\textbf{Text information utilization.} Video-only methods like V2SFlow (28.5\% WER on LRS3) 
face inherent linguistic ambiguity from homovisemes. Our substantially lower error rates 
(17.9\% WER for VTTS TV-CoTemporal on the full LRS3 set, Table~\ref{tab:lrs3_tab_obj})
confirm that the unified decoder effectively uses text to resolve visual
ambiguities---validating that the architectural design properly integrates
linguistic information.

\textbf{Generalization from diverse training.} On full-set LRS3, the directly
comparable LRS3-trained VTTS (TV-CoTemporal) model achieves 17.7\% WER, while
our VoxCeleb2-trained TV-CoTemporal model reaches 17.9\% WER zero-shot
(Table~\ref{tab:lrs3_tab_obj}). Zero-shot transfer without adaptation is the question this work asks; the LRS3-trained row is the maximum-adaptation endpoint for the same variant on the same test set, and the near-parity between the two endpoints (17.9\% vs 17.7\%) is consistent with training on diverse large-scale data transferring across domains.

In Table~\ref{tab:lrs3_tab_obj}, we further evaluate models trained only on
VoxCeleb2. On full-set LRS3, the text-only TTS baseline gives 27.8\% WER,
while adding video with TV-CoTemporal reduces this to 17.9\% WER. Crucially,
the same training data does not guarantee gains for other synchronization
schemes: TV-Global drops to 45.1\%, VT-Global to 38.5\%, and VT-Scaled to
59.6\%. Thus, video conditioning helps only when ordering and position-ID
design are compatible with cross-domain synchronization. 
Because both training corpora are evaluated on the same LRS3 test set, the ordering conclusion is checked under two different training sets: TV-CoTemporal is the better ordering when trained on VoxCeleb2 and evaluated zero-shot (17.9\% vs 59.6\%) and when trained directly on LRS3 (17.7\% vs 28.9\%). VoxCeleb2 PL.v2 contains 1.6k hours from 6k speakers with in-the-wild variability, 
larger and more diverse than the 439h LRS3 training set used by the directly comparable baselines in Table~\ref{tab:lrs3_tab_obj}. All reported results are therefore obtained under noisy conditions: every model trains with random span masking on video, text and speech, and VoxCeleb2 contains in-the-wild, noisy, overlapping speech with variable lighting, pose and motion blur. Zero-shot evaluation on studio-recorded LRS3 additionally measures a noisy-to-clean shift, and that is where the orderings separate most. For transfer-focused analysis with
synchronization scores, we additionally report the introduced 3--45s LRS3
setting in Appendix Table~\ref{tab:lrs3_tab_obj_3to45}, where TV-CoTemporal
reaches 4.8\% WER and 0.22s \alignmetric. The relative ordering remains
consistent with the full-set trends: among VoxCeleb2-trained models,
TV-CoTemporal is strongest, followed by TTS and VT-Global, while VT-Scaled and
TV-Global remain weaker; the LRS3-trained rows are 5.6\% (TV-CoTemporal)
and 6.3\% (VT-Scaled). This ranking is reliable because it repeats across both
evaluation protocols reported in this paper (full-set LRS3 and the 3--45s
transfer-focused setting) and also matches the ordering within the LRS3-trained
pair.
VoiceCraft-Dub* (1.38\% WER) is shown for context only and is not directly
comparable: it builds on a large pretrained VoiceCraft TTS model, uses
Encodec-style discrete speech representations, and applies additional
audio-visual fusion/adaptation during fine-tuning, whereas our VTTS models are
trained from scratch under a unified decoder setup.

Table~\ref{tab:voxceleb2_oldnewdata} shows a comparison of our proposed models and the TTS baseline trained and evaluated on VoxCeleb2. The first two rows are PL.v1 results kept for continuity with earlier versions of the dataset; all remaining rows use PL.v2 and form the main comparison set (including all reported \alignmetric values). All results demonstrate that incorporating video improves both speech content generation and time synchronization. For \alignmetric{} computation, PL.v2 is treated as a ground truth for VoxCeleb2. We evaluate performance using three types of WER. GT WER is computed on original audio using Whisper and serves as an ASR baseline. GT (discrete) WER is computed on audio reconstructed from ground truth discrete speech tokens using our vocoder, representing the lower bound set by quantization and vocoding artifacts. WER is the main evaluation target, measuring how well the model-generated speech aligns with the ground truth transcript. Our objective is to minimize the gap between WER and GT (discrete) WER, as this gap reflects modeling limitations beyond the quantization-vocoder pipeline. Because every model compared here—TTS, all VTTS orderings and LTTS—uses the identical discretization and vocoder, the 10.1\% floor shifts all absolute WERs equally and cannot create or remove differences between variants; the comparative questions this paper asks are therefore unaffected by it. Our best non-global prefix variant sits 2.1 points above the floor (12.2\% vs 10.1\%), so most of the remaining error comes from the representation pipeline rather than the decoder (Appendix~B). 
Improved annotations in PL.v2 lead to consistently better results compared to PL.v1. Furthermore, we compare our full-video VTTS model to a variant using lip crops (LTTS). The VT-Scaled VTTS variant achieves 12.2\% WER, while corresponding LT-Scaled LTTS gives 14.0\% WER, and the TTS baseline yields 14.7\% WER. These results show that using full-face video provides richer information and that our model effectively leverages it without requiring lip cropping.

Our TTS baseline (dMel) uses the same pipeline as the VTTS models by design, so the only difference between the two is the video input. An external TTS system uses a different speech codec and vocoder, and any difference in its numbers could come from the codec rather than from the modality or the ordering. For external context we report reference-only VoxCeleb2 numbers from VDTTS (48.0\% and 25.0\% WER, Table~\ref{tab:vdtts_external_vox}), with the caveat that protocols differ.


\subsection{Ablations}

\vspace{-0.2cm}
\paragraph{Modality Drop.}
Table~\ref{tab:modality_drop} shows the impact of removing individual 
modalities \textit{during evaluation}. Under VT-Scaled, removing text degrades WER from 12.2\% to 74.5\% and removing video from 12.2\% to 46.4\%, 
confirming that both modalities contribute complementary information.
Both degradations are large ($+62.3$ for w/o T and $+34.2$ for w/o V), so the model cannot recover
good transcription quality from a single modality. The degradation is even more 
severe under TV-CoTemporal: removing text yields 112.0\% WER and removing 
video yields 100.7\% WER, i.e., $+97.9$ and $+86.6$ relative to its 14.1 baseline.
Under TV-CoTemporal, both removals are catastrophic, indicating strong joint
dependence on text and video. The gap between 112.0\% and 100.7\% is small
relative to the overall collapse, so we avoid interpreting it as evidence that
one modality dominates the other in this setting. Relative to VT-Scaled, the
larger collapses in TV-CoTemporal are consistent with tighter cross-modal coupling.
The informative part of this ablation is the magnitudes: the degradation is asymmetric under VT-Scaled ($+62.3$ w/o text vs $+34.2$ w/o video), consistent with text carrying the word content, while under TV-CoTemporal the collapse is far more severe ($+97.9$ and $+86.6$). Sensitivity to losing an input is therefore itself ordering-dependent, differing by 35--52 WER points between the two orderings.

\begin{figure}[t!]
    \centering
    \includegraphics[width=0.7\columnwidth]{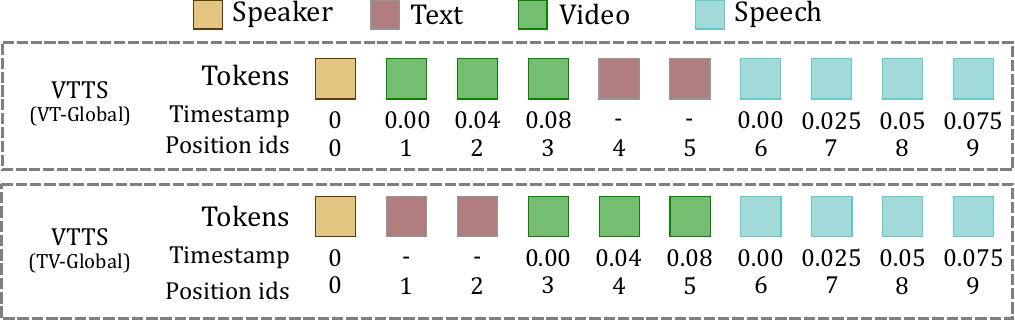}
    \caption{\textbf{Global sequential indexing variants.} Token order, timestamp and position ID for VT-Global (top) and TV-Global (bottom): both keep the prefix orders of Figure~\ref{fig:token_mix} but ignore timestamps and number all tokens by concatenation order. Reading a variant here against its counterpart in Figure~\ref{fig:token_mix} isolates the position-ID computation at fixed token order. Exact metric values are reported in Tables~\ref{tab:voxceleb2_oldnewdata} and~\ref{tab:lrs3_tab_obj}.}
    \label{fig:token_mix_global}
\end{figure}

\vspace{-0.2cm}
\paragraph{Effect of Token Ordering and Position ID Assignment.}
Figure~\ref{fig:token_mix_global} shows the sequence layouts for the global-indexing ablations of the ordered strategies introduced in Section~\ref{sec:multimodal_transformer}.
Across all evaluations, a consistent pattern emerges among non-global prefix variants: VTTS (VT-Scaled) 
achieves stronger in-domain performance (12.2\% WER on VoxCeleb2, 
Table~\ref{tab:voxceleb2_oldnewdata}) and faster convergence 
(Table~\ref{tab:voxceleb2_convergence}), while TV-CoTemporal generalizes 
more robustly to unseen domains (17.9\% vs 59.6\% WER on full-set LRS3 zero-shot, 
Table~\ref{tab:lrs3_tab_obj}), consistent across both objective and 
subjective metrics (Table~\ref{tab:lrs3_mos}). Position ID assignment 
interacts asymmetrically with ordering: under VT ordering, \textit{timestamp-scaled prefix IDs} (VT-Scaled) marginally underperform \textit{global sequential indexing} (VT-Global) on both WER and \alignmetric 
(12.2\% vs 12.1\%, 0.47s vs 0.46s), whereas under TV ordering, \textit{global 
sequential indexing} (TV-Global) improves WER (12.7\% vs 14.1\%) while \textit{co-temporal overlap} (TV-CoTemporal) 
improves \alignmetric (0.44s vs 0.48s). These margins are small in some cases, but they consistently indicate that ordering 
and position ID assignment are not independent design choices. TV-CoTemporal is the stronger variant under transfer. One interpretation is that co-temporal position-ID overlap places video and speech tokens on a shared temporal axis, so tokens that correspond in time receive the same positional index instead of being separated by the prefix offset; we do not test this explanation directly. The in-domain advantage of the video-first variants is likewise reported as an observation. This suggests a practical recommendation: use 
TV-CoTemporal for generalization related applications such as dubbing 
unseen speakers, and VT-Scaled for in-domain applications. We do not claim a single overall-best scheme: the result is a trade-off that holds across objective metrics, subjective MOS and both training sets.

The plain-concatenation variants are competitive in-domain---VT-Global reaches 12.1\% WER, the best value in Table~\ref{tab:voxceleb2_oldnewdata}---but degrade to 38.5\% and 45.1\% on zero-shot LRS3, against 17.9\% for TV-CoTemporal (Table~\ref{tab:lrs3_tab_obj}). It is the gap under domain shift, not in-domain accuracy, that separates the schemes. The interleaved variant matches text-only TTS on WER (14.5\% vs 14.7\%) while improving synchronization (0.49s vs 0.62s \alignmetric, Table~\ref{tab:voxceleb2_oldnewdata}) without surpassing the full-prefix schemes.

\vspace{-0.2cm}
\paragraph{Aggregation of Video Representations.}
Table~\ref{tab:video_agg} shows results of different strategies for spatial aggregation of video representations before inputting into the VTTS (VT-Scaled) decoder. Summation achieves the best WER (12.2\%), but the margin over max pooling (12.4\%) is small, so we treat these two as effectively comparable and choose summation for simplicity. Both significantly outperform more complex strategies like attention (14.5\%) and stacking (14.3\%), suggesting that simple element-wise aggregation might be sufficient for effective video input compression. The ordering of these four strategies is informative on its own: the two parameter-free element-wise operations (summation, max pooling) outperform the two learned ones (attention, stacking) by 1.9--2.3 WER points. A learned aggregator has to discover which of the 256 spatial positions carry articulatory information while the decoder is simultaneously learning cross-modal alignment from scratch, whereas element-wise aggregation preserves the contribution of every position without adding parameters. We therefore use summation for all reported models.

\begin{table*}[t!]
\centering
\begin{minipage}[t]{0.48\textwidth}
\centering
\caption{\textbf{Effect of video aggregation strategies.} Summation performs best, with a small margin over max pooling (12.2\% vs 12.4\%). Both parameter-free element-wise strategies outperform the learned ones (attention, stacking).}
\label{tab:video_agg}
\resizebox{0.7\columnwidth}{!}{
\begin{tabular}{@{}lccccc@{}}\toprule
    \textbf{} & \textbf{Attention} & \textbf{Max} & \textbf{Stacking}  & \textbf{Sum} \\ \midrule
    WER & 14.5 & \underline{12.4} & 14.3 & \textbf{12.2}  \\
\bottomrule
\end{tabular}
}
\end{minipage}
\hfill
\begin{minipage}[t]{0.48\textwidth}
\centering
\caption{\textbf{Modality ablations.} Both modalities are necessary because removing either modality causes large WER increases: VT-Scaled rises from 12.2 to 74.5 (w/o T) or 46.4 (w/o V), and TV-CoTemporal rises from 14.1 to 112.0 (w/o T) or 100.7 (w/o V).}
\label{tab:modality_drop}
\resizebox{\columnwidth}{!}{
\begin{tabular}{@{}lccr@{}}\toprule
    \textbf{Method} & \textbf{GT WER} ($\downarrow$) & \textbf{GT (discrete) WER} ($\downarrow$) & \textbf{WER} ($\downarrow$) \\\midrule
    \multicolumn{4}{@{}l}{\textbf{\textit{VTTS}}} \\
    VT-Scaled & \multirow{3}{*}{2.6 \textpm 0.1} & \multirow{3}{*}{10.1 \textpm 0.2} & \textbf{12.2}  \\
    \qquad w/o T &  &  & 74.5 \\
    \qquad w/o V &  &  & 46.4 \\
    \midrule
    TV-CoTemporal & \multirow{3}{*}{2.6 \textpm 0.1} & \multirow{3}{*}{10.1 \textpm 0.2} & \underline{14.1}  \\
    \qquad w/o T &  &  & 112.0 \\
    \qquad w/o V &  &  & 100.7 \\
\bottomrule
\end{tabular}
}
\end{minipage}
\end{table*}

\begin{figure*}[t!]
    \begin{subfigure}[t]{0.50\textwidth}
    \centering
    \includegraphics[width=0.8\columnwidth]{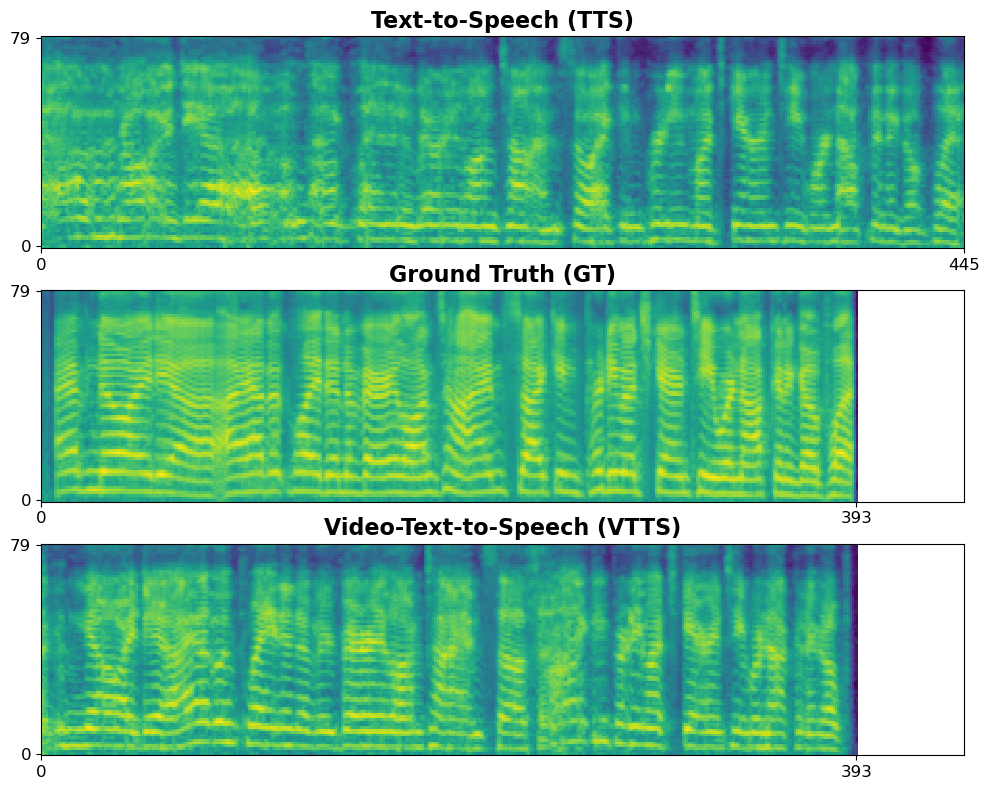}
    \label{fig:qual_mel}
    \end{subfigure}
    \hfill
    \begin{subfigure}[t]{0.48\textwidth}
    \centering
    \includegraphics[width=0.7\columnwidth]{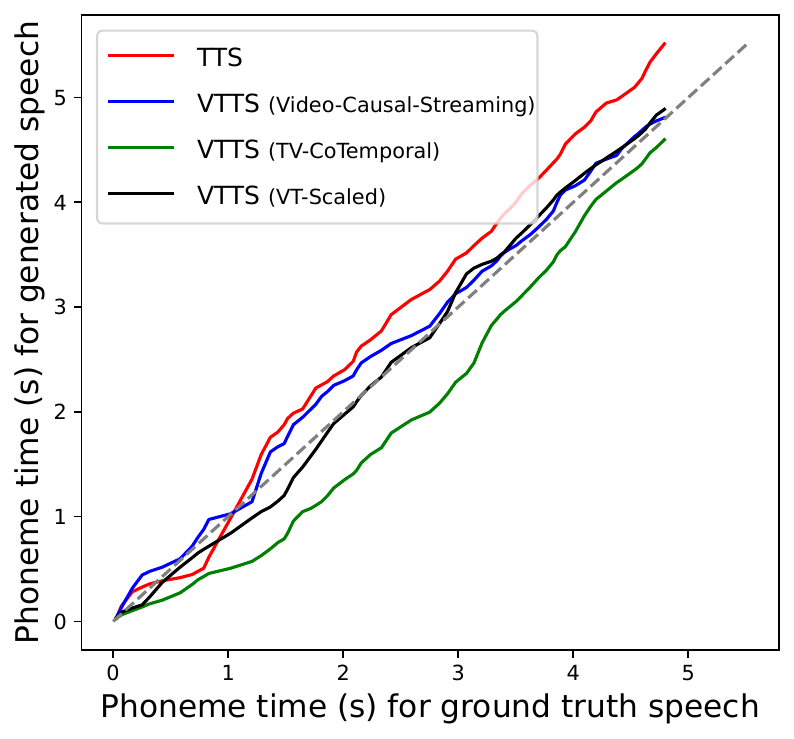}
    \label{fig:metric_example}
    \end{subfigure}
\caption{\textbf{Spectrogram analysis (left):} Log mel-spectrogram comparison for TTS (top), GT (middle), and VTTS (VT-Scaled, bottom). VTTS (VT-Scaled) better matches GT’s timing (393 frames) and energy patterns, unlike TTS which overextends (445 frames). \textbf{Qualitative alignment (right):} \alignmetric visualization of phoneme alignment for the same example. Ground truth and generated phoneme segment centers are plotted on $x$- and $y$-axes respectively, with ideal sync shown as a gray diagonal. VTTS models stay closer to the diagonal than TTS, reflecting better temporal alignment. The per-utterance trajectories shown here are a single example; test-set statistics for the same quantity are in Figure~\ref{fig:metric_pdf} and Table~\ref{tab:voxceleb2_oldnewdata}.}
\label{fig:qual_mel_metric}
\end{figure*}

\vspace{-0.2cm}
\paragraph{Qualitative Results.}
Figure~\ref{fig:qual_mel_metric} shows mel-spectrogram comparisons in which VTTS (VT-Scaled) matches GT's duration and pause locations (393 frames) where TTS overruns (445 frames); further examples, including a failure case, are in the Appendix.

\vspace{-0.2cm}
\paragraph{Training Efficiency.}
Under VT-Scaled, video conditioning improves final performance and accelerates convergence relative to TTS (Table~\ref{tab:voxceleb2_convergence}). At the same 2M-step budget, VTTS (VT-Scaled) reaches 12.2\% WER while TTS is still at 17.3\%, and even after 3M steps TTS only improves to 14.7\%. VTTS (VT-Scaled) has effectively converged by 2M (12.2\% at both 2M and 3M), whereas LTTS (LT-Scaled) continues improving from 14.9\% to 14.0\%, indicating slower convergence for lip-only conditioning. In contrast, TV-CoTemporal at 2M (17.0\%) is close to TTS at 2M, so acceleration is ordering-dependent rather than uniform across variants.

\vspace{-0.2cm}
\paragraph{Statistical reliability.}
All MOS results carry 95\% confidence intervals from bootstrap resampling with 10k iterations (Appendix~\ref{app:subjective_evals}), \alignmetric is reported as mean $\pm$ standard deviation over the test set (Tables~\ref{tab:voxceleb2_oldnewdata} and~\ref{tab:lrs3_tab_obj_3to45}), and GT WER carries $\pm$0.1 (Table~\ref{tab:voxceleb2_oldnewdata}). The conclusions above rest on large margins---17.9\% vs 59.6\% WER for transfer, $+34.2$ to $+97.9$ for modality removal, and 0.44--0.49s vs 0.62s \alignmetric together with non-overlapping MOS intervals (1.98 $\pm$0.15 vs 2.50 $\pm$0.21, Table~\ref{tab:voxceleb2_mos})---while the small margins noted above are treated as ties rather than findings. Robustness here comes from replication rather than reseeding: the ordering ranking repeats across two evaluation protocols, two training sets, and objective and subjective metrics, and each model takes 3--5 days on 8$\times$H100 (Appendix~\ref{app:training_details}).

\begin{table}[t]
\caption{\textbf{Convergence analysis.}~~~Training-iteration comparison shows faster convergence when video modality is used in addition to text. At 2M iterations, VTTS variants are comparable to or better than TTS at the same budget, and VTTS (VT-Scaled) already reaches 12.2\% WER. TTS reaches 14.7\% WER at 3M iterations, while VTTS (VT-Scaled) maintains a smaller gap (+2.1\%) from GT (discrete) WER, indicating more efficient optimization with multimodal conditioning.}
\label{tab:voxceleb2_convergence}
\vspace{0.2cm}
\centering
\resizebox{0.7\columnwidth}{!}{
\begin{tabular}{@{}lcccccc@{}}\toprule
    \textbf{Method} & \textbf{Iterations} & \textbf{GT WER} ($\downarrow$) & \textbf{GT (discrete) WER} ($\downarrow$) & \textbf{WER} ($\downarrow$) \\\midrule
    TTS & 2M & \multirow{5}{*}{2.6 \textpm 0.1} & \multirow{5}{*}{10.1 \textpm 0.2} & 17.3{\tiny{\textcolor{teal}{$\mathord+7.2$}}} \\
    \multicolumn{2}{@{}l}{\textbf{\textit{VTTS}}} & & & \\
    TV-CoTemporal & 2M &  &  & 17.0{\tiny{\textcolor{teal}{$\mathord+6.9$}}} \\
    LTTS (LT-Scaled) & 2M &  &  & \underline{14.9}{\tiny{\textcolor{teal}{$\mathord+4.8$}}} \\
    VT-Scaled & 2M &  &  & \textbf{12.2}{\tiny{\textcolor{teal}{$\mathord+2.1$}}} \\
    \midrule
    TTS & 3M & \multirow{5}{*}{2.6 \textpm 0.1} & \multirow{5}{*}{10.1 \textpm 0.2} & 14.7{\tiny{\textcolor{teal}{$\mathord+4.6$}}} \\
    \multicolumn{2}{@{}l}{\textbf{\textit{VTTS}}} & & & \\
    TV-CoTemporal & 3M &  &  & 14.1{\tiny{\textcolor{teal}{$\mathord+4.0$}}} \\
    LTTS (LT-Scaled) & 3M &  &  & \underline{14.0}{\tiny{\textcolor{teal}{$\mathord+3.9$}}} \\
    VT-Scaled & 3M &  &  & \textbf{12.2}{\tiny{\textcolor{teal}{$\mathord+2.1$}}} \\
\bottomrule
\end{tabular}
}
\end{table}

\contribblock{VoxCeleb2 PL.v2:}{1.6k hours, 6k speakers, 
pseudo-labeled via Demucs + Whisper-large-v2 with quality filtering. 
Enables zero-shot transfer to full-set LRS3 (17.9\% WER, 
Table~\ref{tab:lrs3_tab_obj}) and substantially outperforms directly comparable video-only baselines (V2SFlow: 28.5\%, LipVoicer: 21.4\%; Table~\ref{tab:lrs3_tab_obj}), demonstrating 
that dataset scale and diversity may outweigh strict domain matching.}

\contribblock{Key Findings:}{(i) both modalities are necessary because removing text degrades WER from 12.2\% to 74.5\%, while removing video degrades WER from 12.2\% to 46.4\% (Table~\ref{tab:modality_drop}); (ii) video conditioning improves synchronization over the text-only TTS baseline because \alignmetric drops from 0.62s to 0.47s (Table~\ref{tab:voxceleb2_oldnewdata}); (iii) ordering shows a consistent trade-off among non-global prefix variants because VT-Scaled is stronger in-domain on VoxCeleb2 (12.2\% vs 14.1\% WER), while TV-CoTemporal transfers better zero-shot on full-set LRS3 (17.9\% vs 59.6\% WER) (Tables~\ref{tab:voxceleb2_oldnewdata},~\ref{tab:lrs3_tab_obj}); (iv) under VT-Scaled, video conditioning accelerates convergence because at 2M steps VTTS reaches 12.2\% WER while TTS is at 17.3\%, and even TTS at 3M remains at 14.7\% (Table~\ref{tab:voxceleb2_convergence}).}

\subsection{\alignmetric Analysis} \label{sec:timesync_anl}

Beyond word accuracy, temporal alignment is critical for video-conditioned speech generation. As shown in Table~\ref{tab:voxceleb2_oldnewdata}, video-conditioned models consistently achieve lower \alignmetric error than the text-only TTS baseline. Specifically, VTTS (VT-Scaled) reduces the average phoneme-level timing error from 0.62 ± 0.98 s (TTS) to 0.47 ± 0.63 s, indicating substantially tighter synchronization with the visual stream. Importantly, this improvement is not limited to mean performance: the reduced variance demonstrates that VTTS produces more stable alignment across diverse utterances, whereas TTS exhibits occasional large timing deviations that are perceptually disruptive when overlaid on video.
Comparing different conditioning strategies, both VTTS (Video-Causal-Streaming) and VTTS (TV-CoTemporal) outperform TTS, confirming that access to visual information provides explicit temporal cues unavailable to text-only models. Furthermore, the comparable \alignmetric scores between full-face non-global prefix variants VTTS (0.44–0.47 s) and lip-only LTTS (0.46 s) suggest that lip motion provides the primary temporal signal, with any additional contribution from full-face context depending on ordering rather than being uniform across variants. Overall, these results indicate that \alignmetric is sensitive to the timing differences video conditioning produces, and that explicit video conditioning leads to more precise phoneme-level alignment than text-only synthesis.

\paragraph{Sensitivity and agreement with human judgments.} \alignmetric and human synchronization MOS agree wherever the differences are large enough to interpret. On VoxCeleb2, every video-conditioned variant outperforms TTS on both metrics (0.44–0.49s vs 0.62s TimeSync, Table 4; 2.28–2.50 vs 1.98 synchronization MOS, Table 9). On LRS3, TV-CoTemporal outperforms VT-Scaled on both (0.22s vs 0.32s, Table 10; 3.12 ±0.27 vs 2.35 ±0.22, Table 1, non-overlapping intervals). Within the VoxCeleb2 video-conditioned group the gaps are small on both metrics, and we do not read a ranking into them. The scale is anchored perceptually by the same pairing: the TTS baseline at 0.62s receives synchronization MOS 1.98 while the video-conditioned variants at 0.44–0.49s receive 2.28–2.50, and a bad offset in this data is the TTS tail at 1–2s absolute (Figure 8, right), concentrated in later portions of utterances (Figure 7, right). Three controls bound the metric's sensitivity to its inputs. First, generated and ground truth audio pass through the same forced-alignment pipeline, so a systematic aligner error affects both and largely cancels in the phoneme-center difference \texttt{TimeSync} measures. Second, transcript error is estimated and reported: GT WER is 2.6~$\pm$0.1\% on VoxCeleb2 (Table~4), and LRS3 transcripts are human-annotated. Third, the 3--45s LRS3 setting (Appendix~I, Table~10) exists because we identified and controlled this sensitivity: clips shorter than 3s inflate ASR error and clips longer than 45s accumulate ASR and forced-alignment drift, and the same filter is applied to every system. Alignment and transcript errors are therefore controlled by design, and their size is reported.

\begin{figure}[t]
    \centering
    \includegraphics[width=0.7\columnwidth]{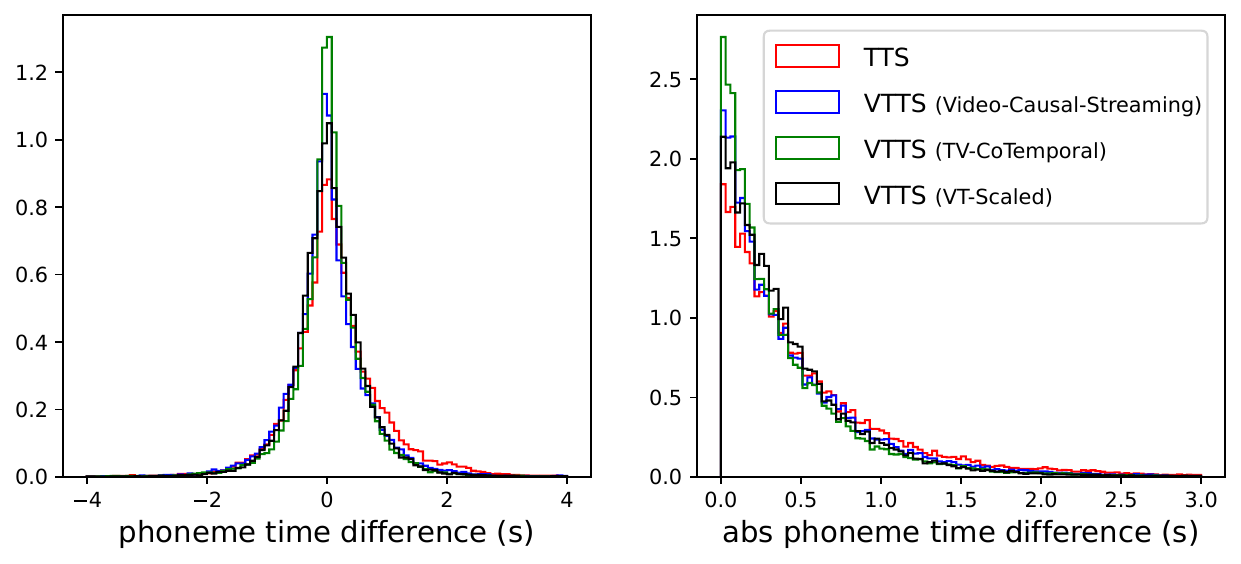}
    \caption{\textbf{Distribution of \alignmetric phoneme timing differences.} 
    We show the difference (left) and absolute difference (right) between 
    ground truth and generated speech phoneme locations (center of phoneme 
    segment) measured in seconds. The $y$-axis is probability density: each panel shows the normalized 
    distribution of phoneme timing differences, so curves are comparable across variants despite 
    differing phoneme counts. The left panel shows symmetric, zero-centered 
    distributions indicating no systematic temporal bias. The right panel 
    reveals that VTTS variants (Video-Causal-Streaming: blue, TV-CoTemporal: green, VT-Scaled: black) concentrate 
    phoneme timing errors within 0-0.5s, while TTS (red) exhibits heavier tails 
    with higher probability of large timing errors (>1.0s). Video-Causal-Streaming appears intermediate between TTS and ordered variants, consistent with 
    more precise timing under video conditioning.}
    \label{fig:metric_pdf}
    \vspace{-0.4cm}
\end{figure}

\paragraph{Temporal Alignment Quality.}
\alignmetric{} can provide deeper insight into temporal alignment quality of the generated speech: we analyze the 
distribution of phoneme-level timing differences between generated and ground 
truth speech. Figure~\ref{fig:metric_pdf} visualizes both the signed difference 
(left) and absolute difference (right) between phoneme center locations in 
generated versus ground truth audio.

The left panel reveals that all models show approximately zero-centered 
distributions, indicating no systematic temporal bias (e.g., consistently 
generating speech too fast or too slow). However, the distribution shapes 
differ significantly: TTS (red) exhibits the widest spread with heavier tails, 
reflecting larger timing errors and inconsistent temporal alignment. In contrast, 
video-conditioned variants show tighter, more concentrated distributions around zero: TV-CoTemporal and VT-Scaled are the sharpest, while Video-Causal-Streaming lies in between full-ordering variants and TTS.

The right panel (absolute differences) more clearly illustrates synchronization 
quality. VTTS models concentrate most phonemes within 0-0.5 seconds of their 
ground truth timing, with probability density rapidly decreasing beyond this 
range. TTS shows notably higher probability mass at larger timing errors (1.0-2.0s), 
indicating more frequent synchronization failures. Video-Causal-Streaming again appears intermediate: better than TTS but less concentrated than TV-CoTemporal/VT-Scaled. This distribution analysis confirms that video conditioning not only improves average \alignmetric scores (Table~\ref{tab:voxceleb2_oldnewdata}: 0.44--0.49s for VTTS variants vs 0.62s for TTS) but also reduces the variance and tail risk of synchronization errors.

Crucially, the combination of a near-zero-centered signed distribution and a
heavy-tailed absolute distribution implies that failures are not primarily
global rate errors; they are intermittent local breakdowns. In other words,
most phonemes are well aligned, but a subset of segments exhibits large
temporal slips. Combined with the per-utterance trajectories in
Figure~\ref{fig:qual_mel_metric} (right), these large deviations are more
common in later portions of an utterance, consistent with accumulated
timing drift rather than uniform misalignment from the beginning.
The complementary qualitative failure mode is that timing can be preserved while spectral quality is not: in the appendix examples (Figures~\ref{fig:qual_mel_suppl_490} and~\ref{fig:metric_example_suppl_490}), VTTS holds closer temporal alignment to the ground truth (165 frames vs 176 for TTS) while both systems fail to reproduce the ground truth's harmonic structure and energy distribution, with degradation most visible in the mid-frequency range.

\vspace{-0.2cm}
\paragraph{VTTS vs. LTTS.} 
We run \alignmetric for LTTS 
on VoxCeleb2 test and obtain 0.46 ±0.61. The matched comparison is against VT-Scaled, which shares its ordering and differs only in using full-face frames rather than lip crops, and the two are effectively tied on this metric (0.47 ±0.63 vs 0.46 ±0.61). Full-face conditioning is ahead of lip-only under TV ordering, where TV-CoTemporal reaches 0.44 ±0.65, so the benefit of the additional facial cues for phoneme timing is ordering-dependent rather than uniform. On WER the matched pair separates more clearly in favor of full-face video (12.2\% vs 14.0\%, Table ~\ref{tab:voxceleb2_oldnewdata}), so the two input types differ more in content accuracy than in timing.

\section{Related Work}

Besides \textit{video-text-to-speech} (VTTS, our focus) setting, speech synthesis from visual input has been explored also in \textit{video-only-to-speech} (V2S) setting. Video-only methods generate speech solely from lip movements, achieving high perceptual quality but facing inherent ambiguity from homophenes: visually identical lip movements for phonetically distinct sounds (e.g., ``bat'' vs. ``pat''). Recent V2S works trained and evaluated on LRS3~\citep{afouras2018lrs3}, such as DiffV2S~\citep{choi2023diffv2s} (word error rate of 39.2\%) and V2SFlow~\citep{choi2025v2sflow} (word error rate of 28.5\%), achieve impressive naturalness but poor intelligibility, as they must resolve linguistic ambiguity from visual information alone. In applications such as movie dubbing, video game localization, audiobook production with video, or accessibility tools, ground-truth text is available alongside video. In these scenarios, \textit{video-text-to-speech} (\task) can leverage both modalities to generate speech that is simultaneously intelligible (from text), temporally synchronized (from video), and expressively aligned with the speaker's facial movements.
Recent \task work includes HPMDubbing~\citep{cong2023learning} and 
StyleDubber~\citep{cong2024styledubber}, which use video and text as inputs 
but have not demonstrated cross-dataset generalization or evaluation on 
large-scale diverse training data like VoxCeleb2~\citep{chung18b_interspeech}.
Also, no prior work has demonstrated practical zero-shot cross-dataset generalization for video-text-to-speech synthesis on large-scale diverse training data.

Speech synthesis models have made rapid progress in recent years, yet they typically focus on text-to-speech or simplified variants of video-conditioned speech generation, e.g., from cropped lip videos~\citep{yemini2024lipvoicer} or generic audio generation~\citep{kondratyuk2024videopoet}. While cropped lip approaches simplify the problem through pretrained lip detectors, they omit important visual cues from the full face that contribute to accurate modeling of phoneme sounds—particularly those involving regions beyond the lips. Furthermore, many prior approaches rely on pretrained ASR~\citep{yemini2024lipvoicer} or follow a pipeline of lip-reading followed by speech generation, which can degrade in noisy, multi-speaker environments where parts of the speech signal may be lost or masked by background interference. To understand how decoder-only architectures learn fine-grained temporal alignment across heterogeneous modalities, we investigate \textit{video-text-to-speech} (VTTS) as a testbed. Unlike prior settings that use single input modalities or specialized architectures, VTTS uniquely combines video (40ms frames), text (sparse tokens), and speech (25ms frames) in a unified decoder. This enables investigation of four key architectural questions: (1) How does explicit text conditioning compare to implicit linguistic knowledge in pretrained encoders? (2) Can unified processing learn temporal alignment without task-specific modules? (3) Does training from scratch on diverse data enable better generalization than domain-specific or pretrained approaches? (4) How should synchronization be measured beyond frame-level metrics to diagnose phoneme-level timing errors? Below, we discuss how existing approaches address related problems and what questions they leave unanswered.

\vspace{-0.2cm}
\paragraph{Text-to-Speech (TTS).}
TTS systems have evolved from early statistical approaches to end-to-end neural methods~\citep{zen2009statistical,kim2020glow,lee2021multi,mehta2024matcha,popov2021grad,ren2019fastspeech,shen2018natural}. Traditional TTS faces challenges with unseen speaker styles, leading to approaches that extract speaker representations from speech data~\citep{chen2021adaspeech,huang2022generspeech,jia2018transfer,lee2022pvae,min2021meta} or incorporate face images to capture visual-acoustic correlations~\citep{goto2020face2speech,lee2023imaginary,wang2022residual}. However, static face images often neglect motion-related factors, leading to inconsistent voice generation when facial expressions vary. Recent unified architectures like VioLA~\citep{wang2023viola} and VOXTLM~\citep{maiti2024voxtlm} attempt to unify speech and text modeling but rely on multi-stage hierarchical processing or lose acoustic detail through discrete content tokens. While TTS excels at generating intelligible speech from text, it lacks visual temporal dynamics, preventing investigation of cross-modal temporal synchronization and how facial expressions influence prosody.

\vspace{-0.2cm}

\paragraph{Lip-to-Speech (LTS).}
LTS aims to reconstruct speech from silent videos of lips—crucial for scenarios with corrupted or missing audio. Early GAN-based approaches like Lip2Wav~\citep{prajwal2020learning}, VCA-GAN~\citep{kim2021lip}, and Lip2Speech~\citep{kim2023lip} demonstrated success on limited vocabulary datasets. Recent work has explored discrete token representations through AV-HuBERT~\citep{shi2022learning}, with ReVISE~\citep{hsu2023revise} integrating HiFi-GAN for improved generation. Diffusion models have also shown promise, with LipVoicer~\citep{yemini2024lipvoicer} achieving 21.4\% WER on LRS3~\citep{afouras2018lrs3}. More recently, V2SFlow~\citep{choi2025v2sflow} addresses video-only synthesis by decomposing speech into content, pitch, and speaker attributes using rectified flow matching, achieving 28.5\% WER on LRS3. While LTS methods do not take explicit text input, modern approaches implicitly encode linguistic knowledge through pretrained audio-visual encoders. For instance, LipVoicer's AV-HuBERT encoder learns to predict masked audio tokens from visual input, providing implicit phonetic structure. However, these approaches face fundamental ambiguity from \textit{homophenes}—visually identical lip movements for phonetically distinct sounds (e.g., ``bat'' vs. ``pat''). Moreover, by focusing primarily on lip movements, they potentially overlook broader visual dynamics from the full face. Without explicit text, LTS cannot answer how explicit linguistic conditioning compares to implicit knowledge, or how models should integrate complementary modalities (text for content, video for timing) in unified architectures.

\vspace{-0.2cm}
\paragraph{Visual Voice Cloning and Dubbing.}
Visual voice cloning~\citep{chen2022v2c,jung2020neural} and dubbing~\citep{cong2023learning,hu2021neural,zhang2024speaker,cong2024styledubber,liu2024m3tts} address related challenges but operate under different constraints. Voice cloning typically aims to replicate a speaker's vocal identity and style using short reference audio-video pairs, often without explicit transcript supervision. Dubbing methods like HPMDubbing~\citep{cong2023learning} and StyleDubber~\citep{cong2024styledubber} replace original speech with new content in different voices or languages, relying on reference speech and specialized modules for duration, pitch, and energy prediction. Recent work has also explored VTTS through transfer learning: VoiceCraft-Dub~\citep{sung2025voicecraft} fine-tunes pretrained VoiceCraft (TTS) with audio-visual fusion layers to achieve 1.38\% WER on LRS3. While this demonstrates effective adaptation of pretrained models, it requires access to large-scale pretraining infrastructure and employs complex Encodec-based representations. While these approaches achieve strong in-domain performance, their reliance on reference audio, pretrained models, or handcrafted alignment modules prevents investigation of whether unified end-to-end architectures can learn robust temporal alignment from scratch on diverse data without such scaffolding.

\section{Limitations}
Further limitations, including representation-ceiling effects from discretization and vocoding, are discussed in Appendix~\ref{app:limitations}.
\section{Conclusion}

We investigate a unified decoder-only transformer for video-text-to-speech that processes visual dynamics, textual content, and speech as discrete tokens in a shared embedding space. Our results show that unified multimodal optimization leverages complementary information: text is important for lexical content, while video provides temporal and expressive cues that improve synchronization, emotional alignment, and intelligibility when combined with text. We also identify an ordering trade-off: video-first ordering gives stronger in-domain performance, while among the timestamp-informed prefix variants text-first ordering transfers more robustly across domains. 
Training on diverse, noisy VoxCeleb2 enables strong zero-shot generalization to full-set LRS3 (17.9\% WER), outperforming prior video-only baselines trained on LRS3 (e.g., LipVoicer: 21.4\%, V2SFlow: 28.5\%) and reaching near-parity with the LRS3-trained TV-CoTemporal counterpart (17.7\%). These results are consistent with the view that diverse large-scale training can reduce reliance on strict domain matching for robust cross-modal representations, though isolating this effect requires targeted ablations. Additionally, our \alignmetric metric provides finer-grained phoneme-level synchronization assessment (0.47$\pm$0.63s vs. 0.62$\pm$0.98s for TTS). These findings establish VTTS as a valuable testbed for understanding how decoder-only architectures should handle heterogeneous temporal modalities.

While the approach used in our systematic analysis achieves strong performance, discretization in both speech and video can remove fine-grained cues: in practice, the speech quantization-vocoder pipeline imposes a measurable reconstruction floor, and video tokenization can suppress subtle articulatory details. The model also occasionally struggles with complex emotional expressions in challenging acoustic conditions. Future work should therefore stay centered on the decoder-only question by studying continuous (or hybrid continuous-discrete) multimodal representations within a single decoder, and by analyzing how ordering and position-ID design affect alignment when the model must process continuous temporal signals directly.

\clearpage
\bibliography{TMLR26/main}

\begin{thebibliography}{71}
\providecommand{\natexlab}[1]{#1}
\providecommand{\url}[1]{\texttt{#1}}
\expandafter\ifx\csname urlstyle\endcsname\relax
  \providecommand{\doi}[1]{doi: #1}\else
  \providecommand{\doi}{doi: \begingroup \urlstyle{rm}\Url}\fi

\bibitem[Achiam et~al.(2023)Achiam, Adler, Agarwal, Ahmad, Akkaya, Aleman,
  Almeida, Altenschmidt, Altman, Anadkat, et~al.]{achiam2023gpt}
Josh Achiam, Steven Adler, Sandhini Agarwal, Lama Ahmad, Ilge Akkaya,
  Florencia~Leoni Aleman, Diogo Almeida, Janko Altenschmidt, Sam Altman,
  Shyamal Anadkat, et~al.
\newblock Gpt-4 technical report.
\newblock \emph{arXiv preprint arXiv:2303.08774}, 2023.

\bibitem[Afouras et~al.(2018)Afouras, Chung, and Zisserman]{afouras2018lrs3}
Triantafyllos Afouras, Joon~Son Chung, and Andrew Zisserman.
\newblock Lrs3-ted: a large-scale dataset for visual speech recognition.
\newblock \emph{arXiv preprint arXiv:1809.00496}, 2018.

\bibitem[Alayrac et~al.(2022)Alayrac, Donahue, Luc, Miech, Barr, Hasson, Lenc,
  Mensch, Millican, Reynolds, et~al.]{alayrac2022flamingo}
Jean-Baptiste Alayrac, Jeff Donahue, Pauline Luc, Antoine Miech, Iain Barr,
  Yana Hasson, Karel Lenc, Arthur Mensch, Katherine Millican, Malcolm Reynolds,
  et~al.
\newblock Flamingo: a visual language model for few-shot learning.
\newblock \emph{Advances in neural information processing systems},
  35:\penalty0 23716--23736, 2022.

\bibitem[Bai et~al.(2022)Bai, Zheng, Chen, Ma, Li, and Huang]{pmlr-v162-bai22d}
He~Bai, Renjie Zheng, Junkun Chen, Mingbo Ma, Xintong Li, and Liang Huang.
\newblock {A}$^3${T}: Alignment-aware acoustic and text pretraining for speech
  synthesis and editing.
\newblock In \emph{Proceedings of the 39th International Conference on Machine
  Learning}, volume 162 of \emph{Proceedings of Machine Learning Research},
  pages 1399--1411. PMLR, 17--23 Jul 2022.
\newblock URL \url{https://proceedings.mlr.press/v162/bai22d.html}.

\bibitem[Bai et~al.(2024)Bai, Likhomanenko, Zhang, Gu, Aldeneh, and
  Jaitly]{bai2024dmel}
He~Bai, Tatiana Likhomanenko, Ruixiang Zhang, Zijin Gu, Zakaria Aldeneh, and
  Navdeep Jaitly.
\newblock dmel: Speech tokenization made simple.
\newblock \emph{arXiv preprint arXiv:2407.15835}, 2024.

\bibitem[Borsos et~al.(2023)Borsos, Marinier, Vincent, Kharitonov, Pietquin,
  Sharifi, Roblek, Teboul, Grangier, Tagliasacchi, et~al.]{borsos2023audiolm}
Zal{\'a}n Borsos, Rapha{\"e}l Marinier, Damien Vincent, Eugene Kharitonov,
  Olivier Pietquin, Matt Sharifi, Dominik Roblek, Olivier Teboul, David
  Grangier, Marco Tagliasacchi, et~al.
\newblock Audiolm: a language modeling approach to audio generation.
\newblock \emph{IEEE/ACM Transactions on Audio, Speech, and Language
  Processing}, 2023.

\bibitem[Carreira et~al.(2018)Carreira, Noland, Banki-Horvath, Hillier, and
  Zisserman]{carreira2018short}
Joao Carreira, Eric Noland, Andras Banki-Horvath, Chloe Hillier, and Andrew
  Zisserman.
\newblock A short note about kinetics-600.
\newblock \emph{arXiv preprint arXiv:1808.01340}, 2018.

\bibitem[Chen et~al.(2019)Chen, Maddox, Duan, and Xu]{chen2019hierarchical}
Lele Chen, Ross~K Maddox, Zhiyao Duan, and Chenliang Xu.
\newblock Hierarchical cross-modal talking face generation with dynamic
  pixel-wise loss.
\newblock In \emph{Proceedings of the IEEE/CVF conference on computer vision
  and pattern recognition}, pages 7832--7841, 2019.

\bibitem[Chen et~al.(2021)Chen, Tan, Li, Liu, Qin, Zhao, and
  Liu]{chen2021adaspeech}
Mingjian Chen, Xu~Tan, Bohan Li, Yanqing Liu, Tao Qin, Sheng Zhao, and Tie-Yan
  Liu.
\newblock Adaspeech: Adaptive text to speech for custom voice.
\newblock \emph{arXiv preprint arXiv:2103.00993}, 2021.

\bibitem[Chen et~al.(2022)Chen, Tan, Qi, Zhou, Li, and Wu]{chen2022v2c}
Qi~Chen, Mingkui Tan, Yuankai Qi, Jiaqiu Zhou, Yuanqing Li, and Qi~Wu.
\newblock V2c: Visual voice cloning.
\newblock In \emph{Proceedings of the IEEE/CVF Conference on Computer Vision
  and Pattern Recognition}, pages 21242--21251, 2022.

\bibitem[Choi et~al.(2023)Choi, Hong, and Ro]{choi2023diffv2s}
Jeongsoo Choi, Joanna Hong, and Yong~Man Ro.
\newblock Diffv2s: Diffusion-based video-to-speech synthesis with vision-guided
  speaker embedding.
\newblock In \emph{Proceedings of the IEEE/CVF International Conference on
  Computer Vision}, pages 7812--7821, 2023.

\bibitem[Choi et~al.(2025)Choi, Kim, Li, Chung, and Liu]{choi2025v2sflow}
Jeongsoo Choi, Ji-Hoon Kim, Jinyu Li, Joon~Son Chung, and Shujie Liu.
\newblock V2sflow: Video-to-speech generation with speech decomposition and
  rectified flow.
\newblock In \emph{ICASSP 2025-2025 IEEE International Conference on Acoustics,
  Speech and Signal Processing (ICASSP)}, pages 1--5. IEEE, 2025.

\bibitem[Chung et~al.(2018)Chung, Nagrani, and Zisserman]{chung18b_interspeech}
Joon~Son Chung, Arsha Nagrani, and Andrew Zisserman.
\newblock {VoxCeleb2: Deep Speaker Recognition}.
\newblock In \emph{Proc. Interspeech 2018}, pages 1086--1090, 2018.
\newblock \doi{10.21437/Interspeech.2018-1929}.

\bibitem[Cong et~al.(2023)Cong, Li, Qi, Zha, Wu, Wang, Jiang, Yang, and
  Huang]{cong2023learning}
Gaoxiang Cong, Liang Li, Yuankai Qi, Zheng-Jun Zha, Qi~Wu, Wenyu Wang, Bin
  Jiang, Ming-Hsuan Yang, and Qingming Huang.
\newblock Learning to dub movies via hierarchical prosody models.
\newblock In \emph{Proceedings of the IEEE/CVF Conference on Computer Vision
  and Pattern Recognition}, pages 14687--14697, 2023.

\bibitem[Cong et~al.(2024)Cong, Qi, Li, Beheshti, Zhang, Hengel, Yang, Yan, and
  Huang]{cong2024styledubber}
Gaoxiang Cong, Yuankai Qi, Liang Li, Amin Beheshti, Zhedong Zhang, Anton
  van~den Hengel, Ming-Hsuan Yang, Chenggang Yan, and Qingming Huang.
\newblock Styledubber: towards multi-scale style learning for movie dubbing.
\newblock \emph{arXiv preprint arXiv:2402.12636}, 2024.

\bibitem[Dai et~al.(2023)Dai, Li, Li, Tiong, Zhao, Wang, Li, Fung, and
  Hoi]{dai2023instructblip}
Wenliang Dai, Junnan Li, Dongxu Li, Anthony Tiong, Junqi Zhao, Weisheng Wang,
  Boyang Li, Pascale~N Fung, and Steven Hoi.
\newblock Instructblip: Towards general-purpose vision-language models with
  instruction tuning.
\newblock \emph{Advances in neural information processing systems},
  36:\penalty0 49250--49267, 2023.

\bibitem[Defossez et~al.(2020)Defossez, Synnaeve, and Adi]{defossez2020real}
Alexandre Defossez, Gabriel Synnaeve, and Yossi Adi.
\newblock Real time speech enhancement in the waveform domain.
\newblock In \emph{Interspeech}, 2020.

\bibitem[Goel et~al.(2025)Goel, Ghosh, Kim, Kumar, Kong, Lee, Yang, Duraiswami,
  Manocha, Valle, et~al.]{goel2025audio}
Arushi Goel, Sreyan Ghosh, Jaehyeon Kim, Sonal Kumar, Zhifeng Kong, Sang-gil
  Lee, Chao-Han~Huck Yang, Ramani Duraiswami, Dinesh Manocha, Rafael Valle,
  et~al.
\newblock Audio flamingo 3: Advancing audio intelligence with fully open large
  audio language models.
\newblock \emph{arXiv preprint arXiv:2507.08128}, 2025.

\bibitem[Goto et~al.(2020)Goto, Onishi, Saito, Tachibana, and
  Mori]{goto2020face2speech}
Shunsuke Goto, Kotaro Onishi, Yuki Saito, Kentaro Tachibana, and Koichiro Mori.
\newblock Face2speech: Towards multi-speaker text-to-speech synthesis using an
  embedding vector predicted from a face image.
\newblock In \emph{INTERSPEECH}, pages 1321--1325, 2020.

\bibitem[Hassid et~al.(2022)Hassid, Tadmor~Ramanovich, Shillingford, Wang, Jia,
  and Remez]{hassid2022vdtts}
Michael Hassid, Michelle Tadmor~Ramanovich, Brendan Shillingford, Miaosen Wang,
  Ye~Jia, and Tal Remez.
\newblock More than words: In-the-wild visually-driven prosody for
  text-to-speech.
\newblock \emph{arXiv preprint arXiv:2111.10139}, 2022.

\bibitem[Hsu et~al.(2023)Hsu, Remez, Shi, Donley, and Adi]{hsu2023revise}
Wei-Ning Hsu, Tal Remez, Bowen Shi, Jacob Donley, and Yossi Adi.
\newblock Revise: Self-supervised speech resynthesis with visual input for
  universal and generalized speech regeneration.
\newblock In \emph{Proceedings of the IEEE/CVF Conference on Computer Vision
  and Pattern Recognition}, pages 18795--18805, 2023.

\bibitem[Hu et~al.(2021)Hu, Tian, Li, Yuping, Wang, and Zhao]{hu2021neural}
Chenxu Hu, Qiao Tian, Tingle Li, Wang Yuping, Yuxuan Wang, and Hang Zhao.
\newblock Neural dubber: Dubbing for videos according to scripts.
\newblock \emph{Advances in neural information processing systems},
  34:\penalty0 16582--16595, 2021.

\bibitem[Huang et~al.(2022)Huang, Ren, Liu, Cui, and
  Zhao]{huang2022generspeech}
Rongjie Huang, Yi~Ren, Jinglin Liu, Chenye Cui, and Zhou Zhao.
\newblock Generspeech: Towards style transfer for generalizable out-of-domain
  text-to-speech.
\newblock \emph{Advances in Neural Information Processing Systems},
  35:\penalty0 10970--10983, 2022.

\bibitem[Jia et~al.(2018)Jia, Zhang, Weiss, Wang, Shen, Ren, Nguyen, Pang,
  Lopez~Moreno, Wu, et~al.]{jia2018transfer}
Ye~Jia, Yu~Zhang, Ron Weiss, Quan Wang, Jonathan Shen, Fei Ren, Patrick Nguyen,
  Ruoming Pang, Ignacio Lopez~Moreno, Yonghui Wu, et~al.
\newblock Transfer learning from speaker verification to multispeaker
  text-to-speech synthesis.
\newblock \emph{Advances in neural information processing systems}, 31, 2018.

\bibitem[Jung and Kim(2020)]{jung2020neural}
Sunghee Jung and Hoirin Kim.
\newblock Neural voice cloning with a few low-quality samples.
\newblock \emph{arXiv preprint arXiv:2006.06940}, 2020.

\bibitem[Kim et~al.(2020)Kim, Kim, Kong, and Yoon]{kim2020glow}
Jaehyeon Kim, Sungwon Kim, Jungil Kong, and Sungroh Yoon.
\newblock Glow-tts: A generative flow for text-to-speech via monotonic
  alignment search.
\newblock \emph{Advances in Neural Information Processing Systems},
  33:\penalty0 8067--8077, 2020.

\bibitem[Kim et~al.(2021)Kim, Hong, and Ro]{kim2021lip}
Minsu Kim, Joanna Hong, and Yong~Man Ro.
\newblock Lip to speech synthesis with visual context attentional gan.
\newblock \emph{Advances in Neural Information Processing Systems},
  34:\penalty0 2758--2770, 2021.

\bibitem[Kim et~al.(2023)Kim, Hong, and Ro]{kim2023lip}
Minsu Kim, Joanna Hong, and Yong~Man Ro.
\newblock Lip-to-speech synthesis in the wild with multi-task learning.
\newblock In \emph{ICASSP 2023-2023 IEEE International Conference on Acoustics,
  Speech and Signal Processing (ICASSP)}, pages 1--5. IEEE, 2023.

\bibitem[Kondratyuk et~al.(2024)Kondratyuk, Yu, Gu, Lezama, Huang, Schindler,
  Hornung, Birodkar, Yan, Chiu, Somandepalli, Akbari, Alon, Cheng, Dillon,
  Gupta, Hahn, Hauth, Hendon, Martinez, Minnen, Sirotenko, Sohn, Yang, Adam,
  Yang, Essa, Wang, Ross, Seybold, and Jiang]{kondratyuk2024videopoet}
Dan Kondratyuk, Lijun Yu, Xiuye Gu, Jose Lezama, Jonathan Huang, Grant
  Schindler, Rachel Hornung, Vighnesh Birodkar, Jimmy Yan, Ming-Chang Chiu,
  Krishna Somandepalli, Hassan Akbari, Yair Alon, Yong Cheng, Joshua~V. Dillon,
  Agrim Gupta, Meera Hahn, Anja Hauth, David Hendon, Alonso Martinez, David
  Minnen, Mikhail Sirotenko, Kihyuk Sohn, Xuan Yang, Hartwig Adam, Ming-Hsuan
  Yang, Irfan Essa, Huisheng Wang, David~A Ross, Bryan Seybold, and Lu~Jiang.
\newblock Videopoet: A large language model for zero-shot video generation.
\newblock In \emph{Forty-first International Conference on Machine Learning},
  2024.
\newblock URL \url{https://openreview.net/forum?id=LRkJwPIDuE}.

\bibitem[Lee et~al.(2022)Lee, Lee, Kim, and Lee]{lee2022pvae}
Ji-Hyun Lee, Sang-Hoon Lee, Ji-Hoon Kim, and Seong-Whan Lee.
\newblock Pvae-tts: Adaptive text-to-speech via progressive style adaptation.
\newblock In \emph{ICASSP 2022-2022 IEEE International Conference on Acoustics,
  Speech and Signal Processing (ICASSP)}, pages 6312--6316. IEEE, 2022.

\bibitem[Lee et~al.(2023)Lee, Chung, and Chung]{lee2023imaginary}
Jiyoung Lee, Joon~Son Chung, and Soo-Whan Chung.
\newblock Imaginary voice: Face-styled diffusion model for text-to-speech.
\newblock In \emph{ICASSP 2023-2023 IEEE International Conference on Acoustics,
  Speech and Signal Processing (ICASSP)}, pages 1--5. IEEE, 2023.

\bibitem[Lee et~al.(2021)Lee, Yoon, Noh, Kim, and Lee]{lee2021multi}
Sang-Hoon Lee, Hyun-Wook Yoon, Hyeong-Rae Noh, Ji-Hoon Kim, and Seong-Whan Lee.
\newblock Multi-spectrogan: High-diversity and high-fidelity spectrogram
  generation with adversarial style combination for speech synthesis.
\newblock In \emph{Proceedings of the AAAI Conference on Artificial
  Intelligence}, volume~35, 2021.

\bibitem[Liu et~al.(2024)Liu, Wei, Qian, Zhang, Chen, and Yin]{liu2024m3tts}
Yan Liu, Li-Fang Wei, Xinyuan Qian, Tian-Hao Zhang, Song-Lu Chen, and Xu-Cheng
  Yin.
\newblock M3tts: Multi-modal text-to-speech of multi-scale style control for
  dubbing.
\newblock \emph{Pattern Recognition Letters}, 179:\penalty0 158--164, 2024.

\bibitem[Lu et~al.(2024)Lu, Clark, Lee, Zhang, Khosla, Marten, Hoiem, and
  Kembhavi]{lu2024unified}
Jiasen Lu, Christopher Clark, Sangho Lee, Zichen Zhang, Savya Khosla, Ryan
  Marten, Derek Hoiem, and Aniruddha Kembhavi.
\newblock Unified-io 2: Scaling autoregressive multimodal models with vision
  language audio and action.
\newblock In \emph{Proceedings of the IEEE/CVF Conference on Computer Vision
  and Pattern Recognition}, pages 26439--26455, 2024.

\bibitem[Ma et~al.(2024)Ma, Zheng, Ye, Li, Gao, Zhang, and
  Chen]{ma2024emotion2vec}
Ziyang Ma, Zhisheng Zheng, Jiaxin Ye, Jinchao Li, Zhifu Gao, Shiliang Zhang,
  and Xie Chen.
\newblock emotion2vec: Self-supervised pre-training for speech emotion
  representation.
\newblock In \emph{Findings of the Association for Computational Linguistics:
  ACL 2024}, pages 15747--15760, 2024.

\bibitem[Maiti et~al.(2024)Maiti, Peng, Choi, Jung, Chang, and
  Watanabe]{maiti2024voxtlm}
Soumi Maiti, Yifan Peng, Shukjae Choi, Jee-weon Jung, Xuankai Chang, and Shinji
  Watanabe.
\newblock Voxtlm: Unified decoder-only models for consolidating speech
  recognition, synthesis and speech, text continuation tasks.
\newblock In \emph{ICASSP 2024-2024 IEEE International Conference on Acoustics,
  Speech and Signal Processing (ICASSP)}, pages 13326--13330. IEEE, 2024.

\bibitem[McKinzie et~al.(2024)McKinzie, Gan, Biard, Dodge, Dufter, Zhang, Shah,
  Du, Peng, Zhang, Weers, Belyi, Singh, Kang, Jain, He, Schwarzer, Gunter,
  Kong, Zhang, Wang, Wang, Du, Lei, Wiseman, Lee, Wang, Pang, Grasch, Toshev,
  and Yang]{mm1}
Brandon McKinzie, Zhe Gan, Jean-Philippe~Fauconnier Biard, Sam Dodge, Philipp
  Dufter, Bowen Zhang, Dhruti Shah, Xianzhi Du, Futang Peng, Haotian Zhang,
  Floris Weers, Anton Belyi, Karanjeet Singh, Doug Kang, Ankur Jain, Hongyu He,
  Max Schwarzer, Tom Gunter, Xiang Kong, Aonan Zhang, Jianyu Wang, Chong Wang,
  Nan Du, Tao Lei, Sam Wiseman, Mark Lee, Zirui Wang, Ruoming Pang, Peter
  Grasch, Alexander Toshev, and Yinfei Yang.
\newblock Mm1: Methods, analysis \& insights from multimodal llm pre-training,
  2024.
\newblock URL \url{https://arxiv.org/abs/2403.09611}.

\bibitem[Mehta et~al.(2024)Mehta, Tu, Beskow, Sz{\'e}kely, and
  Henter]{mehta2024matcha}
Shivam Mehta, Ruibo Tu, Jonas Beskow, {\'E}va Sz{\'e}kely, and Gustav~Eje
  Henter.
\newblock Matcha-tts: A fast tts architecture with conditional flow matching.
\newblock In \emph{ICASSP 2024-2024 IEEE International Conference on Acoustics,
  Speech and Signal Processing (ICASSP)}, pages 11341--11345. IEEE, 2024.

\bibitem[Min et~al.(2021)Min, Lee, Yang, and Hwang]{min2021meta}
Dongchan Min, Dong~Bok Lee, Eunho Yang, and Sung~Ju Hwang.
\newblock Meta-stylespeech: Multi-speaker adaptive text-to-speech generation.
\newblock In \emph{International Conference on Machine Learning}, pages
  7748--7759. PMLR, 2021.

\bibitem[Mira et~al.(2022)Mira, Haliassos, Petridis, Schuller, and
  Pantic]{mira2022svts}
Rodrigo Mira, Alexandros Haliassos, Stavros Petridis, Bj{\"o}rn~W Schuller, and
  Maja Pantic.
\newblock Svts: scalable video-to-speech synthesis.
\newblock \emph{arXiv preprint arXiv:2205.02058}, 2022.

\bibitem[Popov et~al.(2021)Popov, Vovk, Gogoryan, Sadekova, and
  Kudinov]{popov2021grad}
Vadim Popov, Ivan Vovk, Vladimir Gogoryan, Tasnima Sadekova, and Mikhail
  Kudinov.
\newblock Grad-tts: A diffusion probabilistic model for text-to-speech.
\newblock In \emph{International Conference on Machine Learning}, pages
  8599--8608. PMLR, 2021.

\bibitem[Prajwal et~al.(2020{\natexlab{a}})Prajwal, Mukhopadhyay, Namboodiri,
  and Jawahar]{prajwal2020learning}
KR~Prajwal, Rudrabha Mukhopadhyay, Vinay~P Namboodiri, and CV~Jawahar.
\newblock Learning individual speaking styles for accurate lip to speech
  synthesis.
\newblock In \emph{Proceedings of the IEEE/CVF conference on computer vision
  and pattern recognition}, pages 13796--13805, 2020{\natexlab{a}}.

\bibitem[Prajwal et~al.(2020{\natexlab{b}})Prajwal, Mukhopadhyay, Namboodiri,
  and Jawahar]{prajwal2020lip}
KR~Prajwal, Rudrabha Mukhopadhyay, Vinay~P Namboodiri, and CV~Jawahar.
\newblock A lip sync expert is all you need for speech to lip generation in the
  wild.
\newblock In \emph{Proceedings of the 28th ACM international conference on
  multimedia}, pages 484--492, 2020{\natexlab{b}}.

\bibitem[Radford et~al.(2021)Radford, Kim, Hallacy, Ramesh, Goh, Agarwal,
  Sastry, Askell, Mishkin, Clark, et~al.]{radford2021learning}
Alec Radford, Jong~Wook Kim, Chris Hallacy, Aditya Ramesh, Gabriel Goh,
  Sandhini Agarwal, Girish Sastry, Amanda Askell, Pamela Mishkin, Jack Clark,
  et~al.
\newblock Learning transferable visual models from natural language
  supervision.
\newblock In \emph{International conference on machine learning}, pages
  8748--8763. PmLR, 2021.

\bibitem[Radford et~al.(2023)Radford, Kim, Xu, Brockman, McLeavey, and
  Sutskever]{radford2023robust}
Alec Radford, Jong~Wook Kim, Tao Xu, Greg Brockman, Christine McLeavey, and
  Ilya Sutskever.
\newblock Robust speech recognition via large-scale weak supervision.
\newblock In \emph{International conference on machine learning}, pages
  28492--28518. PMLR, 2023.

\bibitem[Ren et~al.(2019)Ren, Ruan, Tan, Qin, Zhao, Zhao, and
  Liu]{ren2019fastspeech}
Yi~Ren, Yangjun Ruan, Xu~Tan, Tao Qin, Sheng Zhao, Zhou Zhao, and Tie-Yan Liu.
\newblock Fastspeech: Fast, robust and controllable text to speech.
\newblock \emph{Advances in neural information processing systems}, 32, 2019.

\bibitem[Sakshi et~al.(2024)Sakshi, Tyagi, Kumar, Seth, Selvakumar, Nieto,
  Duraiswami, Ghosh, and Manocha]{sakshi2024mmau}
S~Sakshi, Utkarsh Tyagi, Sonal Kumar, Ashish Seth, Ramaneswaran Selvakumar,
  Oriol Nieto, Ramani Duraiswami, Sreyan Ghosh, and Dinesh Manocha.
\newblock Mmau: A massive multi-task audio understanding and reasoning
  benchmark.
\newblock \emph{arXiv preprint arXiv:2410.19168}, 2024.

\bibitem[Shen et~al.(2018)Shen, Pang, Weiss, Schuster, Jaitly, Yang, Chen,
  Zhang, Wang, Skerrv-Ryan, et~al.]{shen2018natural}
Jonathan Shen, Ruoming Pang, Ron~J Weiss, Mike Schuster, Navdeep Jaitly,
  Zongheng Yang, Zhifeng Chen, Yu~Zhang, Yuxuan Wang, Rj~Skerrv-Ryan, et~al.
\newblock Natural tts synthesis by conditioning wavenet on mel spectrogram
  predictions.
\newblock In \emph{2018 IEEE international conference on acoustics, speech and
  signal processing (ICASSP)}, pages 4779--4783. IEEE, 2018.

\bibitem[Shi et~al.(2022)Shi, Mohamed, and Hsu]{shi2022learning}
Bowen Shi, Abdelrahman Mohamed, and Wei-Ning Hsu.
\newblock Learning lip-based audio-visual speaker embeddings with av-hubert.
\newblock \emph{arXiv preprint arXiv:2205.07180}, 2022.

\bibitem[Su et~al.(2024)Su, Ahmed, Lu, Pan, Bo, and Liu]{su2024roformer}
Jianlin Su, Murtadha Ahmed, Yu~Lu, Shengfeng Pan, Wen Bo, and Yunfeng Liu.
\newblock Roformer: Enhanced transformer with rotary position embedding.
\newblock \emph{Neurocomputing}, 568:\penalty0 127063, 2024.

\bibitem[Su et~al.(2023)Su, Lan, Li, Xu, Wang, and Cai]{su2023pandagpt}
Yixuan Su, Tian Lan, Huayang Li, Jialu Xu, Yan Wang, and Deng Cai.
\newblock Pandagpt: One model to instruction-follow them all.
\newblock In \emph{Proceedings of the 1st Workshop on Taming Large Language
  Models: Controllability in the era of Interactive Assistants!}, pages 11--23,
  2023.

\bibitem[Sung-Bin et~al.(2025)Sung-Bin, Choi, Peng, Chung, Oh, and
  Harwath]{sung2025voicecraft}
Kim Sung-Bin, Jeongsoo Choi, Puyuan Peng, Joon~Son Chung, Tae-Hyun Oh, and
  David Harwath.
\newblock Voicecraft-dub: Automated video dubbing with neural codec language
  models.
\newblock \emph{arXiv preprint arXiv:2504.02386}, 2025.

\bibitem[Team et~al.(2024)Team, Georgiev, Lei, Burnell, Bai, Gulati, Tanzer,
  Vincent, Pan, Wang, et~al.]{team2024gemini}
Gemini Team, Petko Georgiev, Ving~Ian Lei, Ryan Burnell, Libin Bai, Anmol
  Gulati, Garrett Tanzer, Damien Vincent, Zhufeng Pan, Shibo Wang, et~al.
\newblock Gemini 1.5: Unlocking multimodal understanding across millions of
  tokens of context.
\newblock \emph{arXiv preprint arXiv:2403.05530}, 2024.

\bibitem[Tong et~al.(2022)Tong, Song, Wang, and Wang]{tong2022videomae}
Zhan Tong, Yibing Song, Jue Wang, and Limin Wang.
\newblock Videomae: Masked autoencoders are data-efficient learners for
  self-supervised video pre-training.
\newblock \emph{Advances in neural information processing systems},
  35:\penalty0 10078--10093, 2022.

\bibitem[Touvron et~al.(2023)Touvron, Lavril, Izacard, Martinet, Lachaux,
  Lacroix, Rozi{\`e}re, Goyal, Hambro, Azhar, et~al.]{touvron2023llama}
Hugo Touvron, Thibaut Lavril, Gautier Izacard, Xavier Martinet, Marie-Anne
  Lachaux, Timoth{\'e}e Lacroix, Baptiste Rozi{\`e}re, Naman Goyal, Eric
  Hambro, Faisal Azhar, et~al.
\newblock Llama: Open and efficient foundation language models.
\newblock \emph{arXiv preprint arXiv:2302.13971}, 2023.

\bibitem[Unterthiner et~al.(2018)Unterthiner, Van~Steenkiste, Kurach, Marinier,
  Michalski, and Gelly]{unterthiner2018towards}
Thomas Unterthiner, Sjoerd Van~Steenkiste, Karol Kurach, Raphael Marinier,
  Marcin Michalski, and Sylvain Gelly.
\newblock Towards accurate generative models of video: A new metric \&
  challenges.
\newblock \emph{arXiv preprint arXiv:1812.01717}, 2018.

\bibitem[Variani et~al.(2014)Variani, Lei, McDermott, Moreno, and
  Gonzalez-Dominguez]{variani2014deep}
Ehsan Variani, Xin Lei, Erik McDermott, Ignacio~Lopez Moreno, and Javier
  Gonzalez-Dominguez.
\newblock Deep neural networks for small footprint text-dependent speaker
  verification.
\newblock In \emph{2014 IEEE international conference on acoustics, speech and
  signal processing (ICASSP)}, pages 4052--4056. IEEE, 2014.

\bibitem[Wang et~al.(2022)Wang, Wang, Hu, Li, Fang, and Liu]{wang2022residual}
Jianrong Wang, Zixuan Wang, Xiaosheng Hu, Xuewei Li, Qiang Fang, and Li~Liu.
\newblock Residual-guided personalized speech synthesis based on face image.
\newblock In \emph{ICASSP 2022-2022 IEEE International Conference on Acoustics,
  Speech and Signal Processing (ICASSP)}, pages 4743--4747. IEEE, 2022.

\bibitem[Wang et~al.(2023)Wang, Zhou, Zhang, Wu, Liu, Gaur, Chen, Li, and
  Wei]{wang2023viola}
Tianrui Wang, Long Zhou, Ziqiang Zhang, Yu~Wu, Shujie Liu, Yashesh Gaur, Zhuo
  Chen, Jinyu Li, and Furu Wei.
\newblock Viola: Unified codec language models for speech recognition,
  synthesis, and translation.
\newblock \emph{arXiv preprint arXiv:2305.16107}, 2023.

\bibitem[Wu et~al.(2024)Wu, Fei, Qu, Ji, and Chua]{wu2024next}
Shengqiong Wu, Hao Fei, Leigang Qu, Wei Ji, and Tat-Seng Chua.
\newblock Next-gpt: Any-to-any multimodal llm.
\newblock In \emph{Forty-first International Conference on Machine Learning},
  2024.

\bibitem[Xu et~al.(2025{\natexlab{a}})Xu, Guo, He, Hu, He, Bai, Chen, Wang,
  Fan, Dang, Zhang, Wang, Chu, and Lin]{xu2025qwen25omnitechnicalreport}
Jin Xu, Zhifang Guo, Jinzheng He, Hangrui Hu, Ting He, Shuai Bai, Keqin Chen,
  Jialin Wang, Yang Fan, Kai Dang, Bin Zhang, Xiong Wang, Yunfei Chu, and
  Junyang Lin.
\newblock Qwen2.5-omni technical report, 2025{\natexlab{a}}.
\newblock URL \url{https://arxiv.org/abs/2503.20215}.

\bibitem[Xu et~al.(2025{\natexlab{b}})Xu, Guo, He, Hu, He, Bai, Chen, Wang,
  Fan, Dang, et~al.]{xu2025qwen2}
Jin Xu, Zhifang Guo, Jinzheng He, Hangrui Hu, Ting He, Shuai Bai, Keqin Chen,
  Jialin Wang, Yang Fan, Kai Dang, et~al.
\newblock Qwen2. 5-omni technical report.
\newblock \emph{arXiv preprint arXiv:2503.20215}, 2025{\natexlab{b}}.

\bibitem[Yamamoto et~al.(2020)Yamamoto, Song, and Kim]{yamamoto2020parallel}
Ryuichi Yamamoto, Eunwoo Song, and Jae-Min Kim.
\newblock Parallel wavegan: A fast waveform generation model based on
  generative adversarial networks with multi-resolution spectrogram.
\newblock In \emph{ICASSP 2020-2020 IEEE International Conference on Acoustics,
  Speech and Signal Processing (ICASSP)}, pages 6199--6203. IEEE, 2020.

\bibitem[Yan et~al.(2021)Yan, Zhang, Abbeel, and Srinivas]{yan2021videogpt}
Wilson Yan, Yunzhi Zhang, Pieter Abbeel, and Aravind Srinivas.
\newblock Videogpt: Video generation using vq-vae and transformers.
\newblock \emph{arXiv preprint arXiv:2104.10157}, 2021.

\bibitem[Yang et~al.(2025)Yang, Li, Yang, Zhang, Hui, Zheng, Yu, Gao, Huang,
  Lv, et~al.]{yang2025qwen3}
An~Yang, Anfeng Li, Baosong Yang, Beichen Zhang, Binyuan Hui, Bo~Zheng, Bowen
  Yu, Chang Gao, Chengen Huang, Chenxu Lv, et~al.
\newblock Qwen3 technical report.
\newblock \emph{arXiv preprint arXiv:2505.09388}, 2025.

\bibitem[Yemini et~al.(2024)Yemini, Shamsian, Bracha, Gannot, and
  Fetaya]{yemini2024lipvoicer}
Yochai Yemini, Aviv Shamsian, Lior Bracha, Sharon Gannot, and Ethan Fetaya.
\newblock Lipvoicer: Generating speech from silent videos guided by lip
  reading.
\newblock In \emph{The Twelfth International Conference on Learning
  Representations}, 2024.
\newblock URL \url{https://openreview.net/forum?id=ZZCPSC5OgD}.

\bibitem[Young et~al.(2002)Young, Evermann, Gales, Hain, Kershaw, Liu, Moore,
  Odell, Ollason, Povey, et~al.]{young2002htk}
Steve Young, Gunnar Evermann, Mark Gales, Thomas Hain, Dan Kershaw, Xunying
  Liu, Gareth Moore, Julian Odell, Dave Ollason, Dan Povey, et~al.
\newblock The htk book.
\newblock \emph{Cambridge university engineering department}, 3\penalty0
  (175):\penalty0 12, 2002.

\bibitem[Yu et~al.(2022)Yu, Xu, Koh, Luong, Baid, Wang, Vasudevan, Ku, Yang,
  Ayan, Hutchinson, Han, Parekh, Li, Zhang, Baldridge, and Wu]{parti}
Jiahui Yu, Yuanzhong Xu, Jing~Yu Koh, Thang Luong, Gunjan Baid, Zirui Wang,
  Vijay Vasudevan, Alexander Ku, Yinfei Yang, Burcu~Karagol Ayan, Ben
  Hutchinson, Wei Han, Zarana Parekh, Xin Li, Han Zhang, Jason Baldridge, and
  Yonghui Wu.
\newblock Scaling autoregressive models for content-rich text-to-image
  generation, 2022.
\newblock URL \url{https://arxiv.org/abs/2206.10789}.

\bibitem[Zen et~al.(2009)Zen, Tokuda, and Black]{zen2009statistical}
Heiga Zen, Keiichi Tokuda, and Alan~W Black.
\newblock Statistical parametric speech synthesis.
\newblock \emph{speech communication}, 51\penalty0 (11):\penalty0 1039--1064,
  2009.

\bibitem[Zhan et~al.(2024)Zhan, Dai, Ye, Zhou, Zhang, Liu, Zhang, Yuan, Zhang,
  Li, et~al.]{zhan2024anygpt}
Jun Zhan, Junqi Dai, Jiasheng Ye, Yunhua Zhou, Dong Zhang, Zhigeng Liu, Xin
  Zhang, Ruibin Yuan, Ge~Zhang, Linyang Li, et~al.
\newblock Anygpt: Unified multimodal llm with discrete sequence modeling.
\newblock In \emph{Proceedings of the 62nd Annual Meeting of the Association
  for Computational Linguistics (Volume 1: Long Papers)}, pages 9637--9662,
  2024.

\bibitem[Zhang et~al.(2024)Zhang, Li, Cong, Yin, Gao, Yan, Hengel, and
  Qi]{zhang2024speaker}
Zhedong Zhang, Liang Li, Gaoxiang Cong, Haibing Yin, Yuhan Gao, Chenggang Yan,
  Anton van~den Hengel, and Yuankai Qi.
\newblock From speaker to dubber: movie dubbing with prosody and duration
  consistency learning.
\newblock In \emph{Proceedings of the 32nd ACM International Conference on
  Multimedia}, pages 7523--7532, 2024.

\end{thebibliography}
\bibliographystyle{plainnat}

\clearpage
\appendix

\section*{Appendix}

\section{Ethics Discussion}
\label{app:ethic}

The advancement of speech technologies brings great potential but also significant ethical challenges that must not be overlooked. 
While we aim to create techniques that improve conditional speech synthesis for multimodal settings, it is vital to address risks proactively and promote awareness to guide responsible innovation at different levels: from researchers to the end-users. As such, we highlight several key challenges:
\begin{itemize}
    \item \textbf{Dual-use risk}~~~Improvements in audio-visual synchronization, including the ordering and position-ID findings reported here, could make synthesized talking-head video harder to identify from timing errors: impersonation, voice spoofing, unauthorized dubbing and misleading audio-visual content all become more plausible as timing artifacts diminish.
    
    \item \textbf{Deployment-side defenses}~~~Two are practical today: watermarking generated speech, by inserting detectable markers at synthesis time, and detectors trained to identify synthetic speech. Neither is sufficient alone.
    
    \item \textbf{Consent and governance}~~~Downstream applications using voice and face data should require consent from the individuals whose voice or likeness is synthesized, and datasets of face and voice data warrant governance beyond a research license provenance, retention and permitted-use terms.
    
    \item \textbf{VoxCeleb2 specifically}~~~We use VoxCeleb2 under its research license (Appendix C). Public availability does not, by itself, resolve consent for synthesis research involving faces and voices, and we do not treat it as doing so; this is part of the rationale for withholding trained models. We obtained consent from the individual shown in the example video frames for the use of these images in this paper.
    
    \item \textbf{Safeguards already in effect}~~~We do not release pretrained models, for privacy, safety and misuse reasons (Appendix ~\ref{app:repro}); released material is limited to code and the PL.v2 transcriptions. Phoneme-level timing analysis of the kind TimeSync performs may also serve the defensive side, as a diagnostic for synchronization inconsistencies in manipulated media.

    \item \textbf{Accessibility and inclusivity}~~~While we are working with English-only data for the proof of concept, extending the speech technologies for diverse populations and existing spoken languages should be a top priority in the community.
    
    \item \textbf{Transparency and accountability}~~~Detailed documentation, limitations, analysis of failure cases, and reproducibility are essential for promoting transparency and informed usage. Responsibility in development and deployment should remain a cornerstone in the community.
\end{itemize}

\section{Limitations}\label{app:limitations}

While we made the best effort to tune TTS baseline, there is always a possibility we missed some details. 
Due to optimization issues when both modalities, video and text, are inputted into the model, we first found best hyper-parameters for our VTTS models so that the models can converge. Later, the same hyper-parameters are used for the TTS baseline by excluding video from the input into the model.
However, across our in-domain VoxCeleb2 experiments and hyper-parameter tuning, video-conditioned models consistently outperform the TTS baseline, while under cross-domain transfer the benefit depends strongly on token ordering and position-ID design.


We did not train larger models (>300M parameters), did not use larger datasets (>1.6k hours) or pre-trained models, and leave this as a future work. The same applies to sequence designs that mix position-ID schemes within a single time chunk computing IDs from timestamps for some tokens and from sequence order for their neighbors which the per-token parameterization of Section 2.3 permits but which we do not evaluate.

GT (discrete) WER of 10.1\% defines an empirical floor for models using our current discretization pipeline, since even reconstructed ground-truth tokens incur this error. Our best VoxCeleb2 model among non-global prefix variants (VTTS (VT-Scaled)) reaches 12.2\% WER, only 2.1 absolute points above this floor (VT-Global reaches 12.1\%). This indicates limited remaining headroom within the current representation pipeline and suggests that further gains increasingly depend on improving discretization and reconstruction quality, not only decoder modeling. VoxCeleb2 contains noisy and overlapping speech, which degrades mel-based discretization and vocoder reconstruction. Because the vocoder is trained independently and not optimized for in-the-wild speech, this floor remains a core limitation; future work may reduce it using neural codecs or more noise-robust vocoders.


\section{Data, Code, Reproducibility}\label{app:repro}
We made the best effort to use publicly available data and official implementations (e.g. VQ-VAE for video representations). All data we used are under permissive license for research. We do our best to provide all details and steps in the main text and in Appendix. 
We release all code and the PL.v2 transcriptions for VoxCeleb2 data.

We do not plan to open-source any pre-trained models for sake of privacy, safety and misuse. Withholding trained weights affects reproducibility, and we mitigate this on the release side. We release all code, including the data-processing and evaluation scripts, together with the PL.v2 transcriptions; all data used are publicly available under permissive research licenses; and Appendix ~\ref{app:training_details} gives the complete training recipe architecture ($~250$M parameters, $D'$ = 768, 36 layers), optimizer, learning-rate schedule, masking strategy, batch construction and compute budget (8×H100, 3–5 days) sufficient to retrain every reported model from scratch. Only the trained weights are withheld, for the reasons discussed in Appendix ~\ref{app:ethic}.

\section{Sequence-Design Variants}\label{app:variants}
 
The five variants studied in Section~\ref{sec:multimodal_transformer} are summarized in
Table~\ref{tab:variants}.
 
\begin{table}[t]
\centering
\small
\footnotesize
\setlength{\tabcolsep}{5pt}
\caption{\textbf{Sequence-design variants.} Each variant is fully specified by the three per-token
quantities of Figure~\ref{fig:token_mix}: token order, timestamp, and how the position ID is
computed. A speaker token precedes all modalities in every variant and is omitted from the
table. All variants use full-prefix conditioning-every conditioning token precedes the speech
tokens-except Video-Causal-Streaming, in which video and speech are interleaved in temporal
order so that speech at time $t$ attends only to video at $t' \le t$. Order: T, V, S denote text, video and speech tokens; (I) marks interleaving; VC-Streaming abbreviates Video-Causal-Streaming. Shared: whether video and speech tokens share position IDs. Causal: "prefix" means all conditioning tokens precede speech.}
\label{tab:variants}
\begin{tabular}{@{}llllll@{}}
\toprule
Variant & Order & Position IDs & Shared & Causal & Intended setting \\
\midrule
TV-CoTemporal          & T$\to$V$\to$S          & timestamp, shared axis & yes & prefix          & transfer; dubbing unseen speakers \\
VT-Scaled              & V$\to$T$\to$S          & timestamp-scaled video & no  & prefix          & in-domain synthesis \\
VC-Streaming & T$\to$V/S (I)  & temporal order         & yes & video $t'\le t$ & streaming, low latency \\
TV-Global              & T$\to$V$\to$S          & concatenation order    & no  & prefix          & ID ablation \\
VT-Global              & V$\to$T$\to$S          & concatenation order    & no  & prefix          & ID ablation \\
\bottomrule
\end{tabular}
\end{table}

\section{Video Reconstruction} \label{app:video_recon}

Although the VQ-VAE model used to extract video representations is pre-trained on the general videos, we found it reconstructs speakers videos with sufficient quality to preserve necessary spatial information.
To evaluate video reconstruction quality, we employ the Fr\'{e}chet Video Distance (FVD)~\cite{unterthiner2018towards}, specifically the FVD$_{16}$ variant that assesses quality over 16-frames window. The FVD scores are computed using an I3D model trained on Kinetics-400, providing a standardized measure of video quality across different temporal scales.
The FVD metric is 86.2 at resolution 64x64.
\textit{Thus, we do not finetune the model further on videos of talking people and use it as is.}

\section{Video Representation} \label{app:video_rep}

To obtain a frame-level embedding suitable for autoregressive modeling, we aggregate the spatial grid of quantized video embeddings into a single vector $\mathbf{z}_t^v \in \mathbb{R}^{D'}$ per frame. As shown in Figure 2 (main paper), each frame is first discretized using a pretrained VQ-VAE encoder, resulting in a $H' \times W'$ grid of tokens, where each token is embedded via a learnable table $\mathbf{E}^v$. Since our unified decoder processes one token per time step, we aggregate the $H' \times W'$ spatial embeddings into a single vector before feeding into the decoder model. Below, we outline the aggregation strategies we explored:

\textbf{Attention:} having learnable $Q, K, V\in\mathbb{R}^{D'\times D'}$ and $\text{attn}_{h, w}=\text{softmax}_{h,w}(Q\mathbf{e}^v_{t,1,1}, K\mathbf{e}^v_{t,h,w})$ we compute $\mathbf{z}_t^v = \frac{1}{\sqrt{D'}}\sum_{h=1}^{H'} \sum_{w=1}^{W'} \text{attn}_{h, w} \,\,\, V\mathbf{e}^v_{t,h,w}$;

\textbf{Summation:} $\mathbf{z}_t^v = \sum_{h=1}^{H'} \sum_{w=1}^{W'} \mathbf{e}^v_{t,h,w}$; 

\textbf{Mean pooling:} $\mathbf{z}_t^v = \frac{1}{H'W'}\sum_{h=1}^{H'} \sum_{w=1}^{W'} \mathbf{e}^v_{t,h,w}$; 

\textbf{Max pooling:} $\mathbf{z}_t^v =\max_{(h, w)}\mathbf{e}^v_{t,h,w}$;

\textbf{Stacking:} stack embeddings and then project it via a learnable linear layer $\mathbf{L}^v(\cdot): \mathbb{R}^{H'W'D'} \to \mathbb{R}^{D'}$,
$\mathbf{z}_t^v = \mathbf{L}^v([\mathbf{e}^v_{t,1,1}, \mathbf{e}^v_{t,1,2}, \dots, \mathbf{e}^v_{t,1,W'}, \mathbf{e}^v_{t,2,1}, \dots, \mathbf{e}^v_{t,H',W'}])$.

We choose VQ-VAE over discriminative video representations (e.g., CLIP, AV-HuBERT, VideoMAE) because VTTS requires preserving fine-grained spatial and temporal facial dynamics rather than global semantic alignment. Discriminative encoders are optimized for semantic matching or classification objectives and often discard spatial locality, whereas VQ-VAE retains a grid of discrete visual tokens encoding mouth shape, jaw motion, and facial muscle activations. These cues are critical for modeling phoneme timing, articulation strength, and emotional intensity. While we do not explicitly supervise emotion, our human evaluation results (Table~\ref{tab:voxceleb_mos_emotion}) indicate that these visual tokens implicitly encode facial expressions relevant to speech prosody and emotional delivery.

\section{Evaluation Metrics}

\textbf{Word Error Rate (WER)}~~~We use Whisper-large v2 via open-source code \url{https://github.com/m-bain/whisperX} to transcribe generated speech. The latter is compared to the ground truth transcription (PL.v2 is treated as a ground truth for VoxCeleb2) to compute WER. 
\begin{figure*}[t!]
    \centering
    \includegraphics[width=0.9\textwidth]{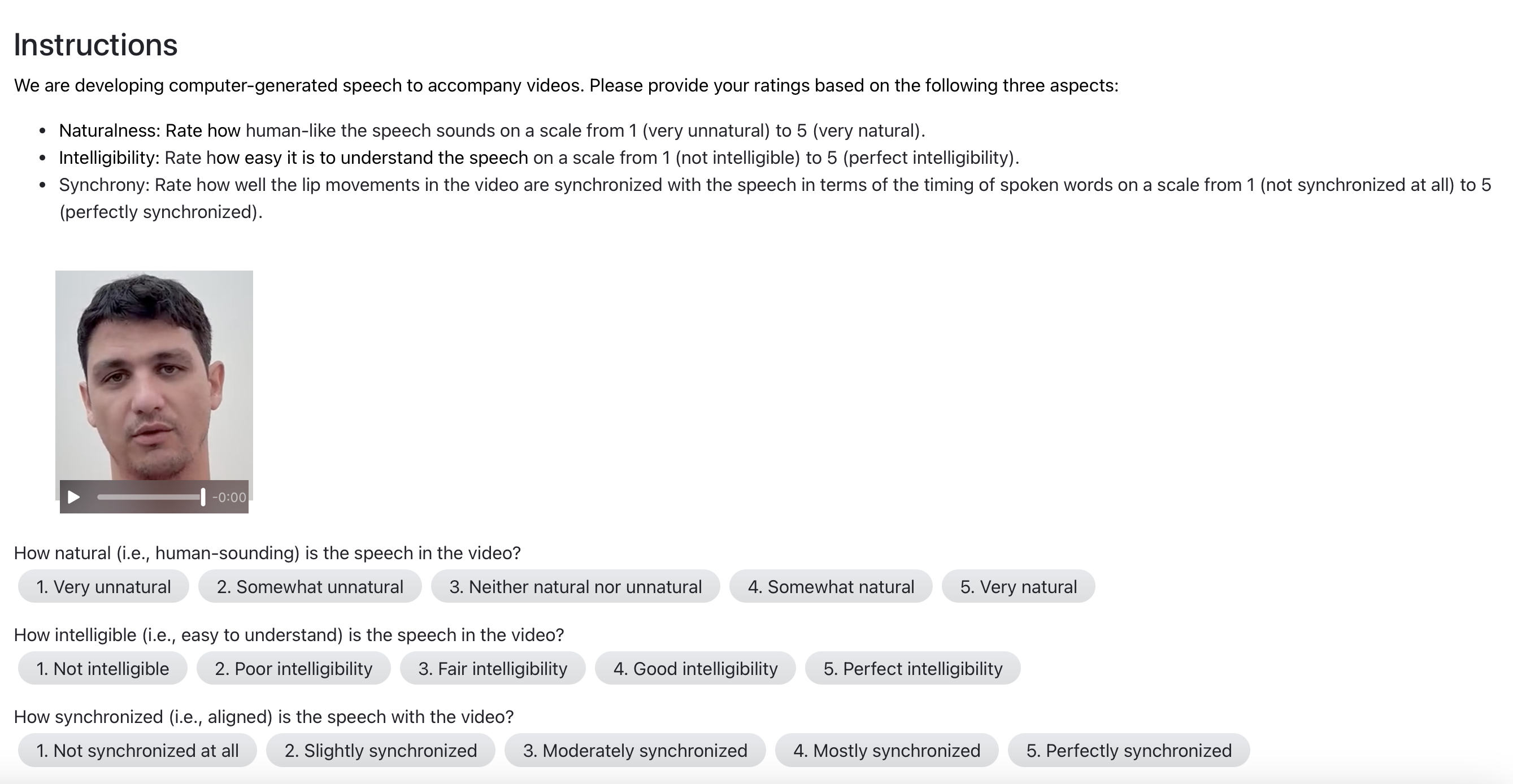}
    \caption{\textbf{Human evaluation.} Task description for the crowd-sourced raters to evaluate intelligibility, naturalness and synchronization of the ground truth or generated speech: speech is overlayed with the video and they are played together for the raters.}
    \label{fig:mos_eval_setup}
    \vspace{-0.4cm}
\end{figure*}
\begin{figure*}[t!]
    \centering
    \includegraphics[width=0.9\textwidth]{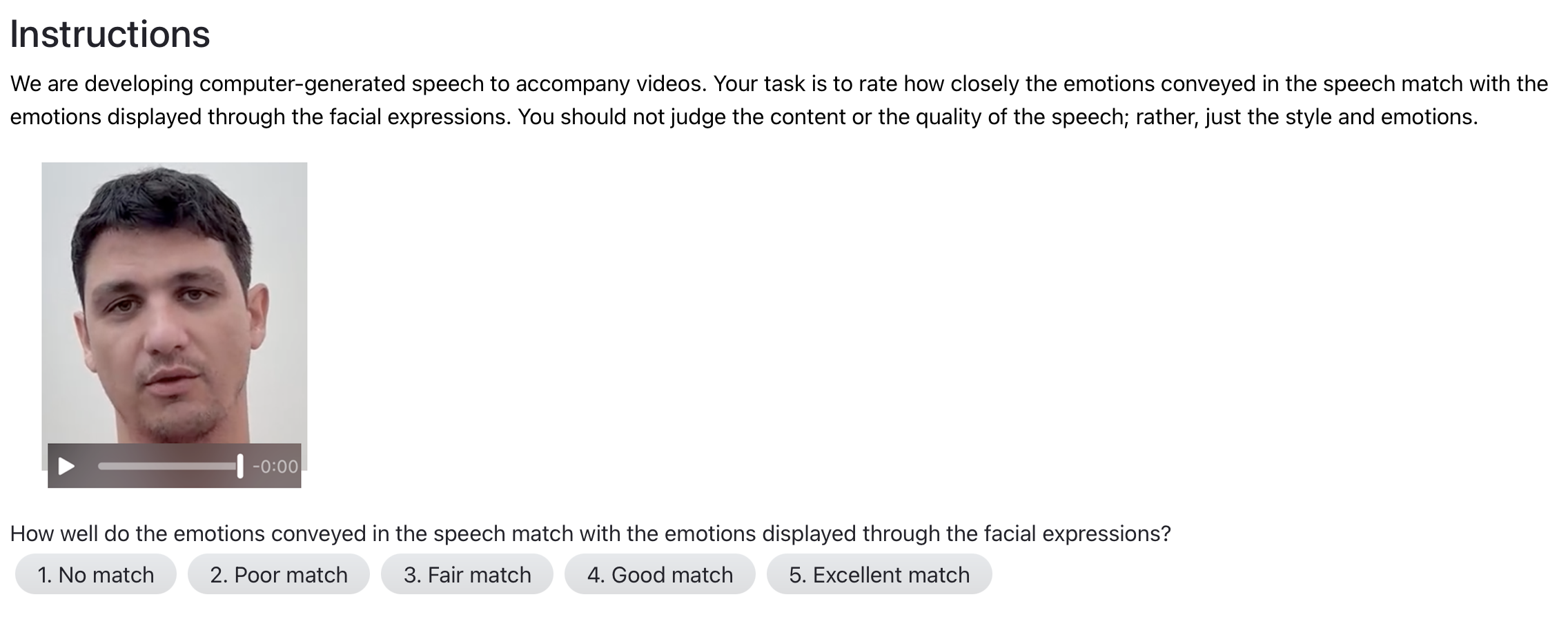}
    \caption{\textbf{Human evaluation.} Task description for the crowd-sourced raters to evaluate correspondence between facial expressions and emotions in speech for ground truth and generated speech: speech is overlayed with the video and they are played together for the raters.}
    \label{fig:mos_eval_setup_emotion}
    \vspace{-0.4cm}
\end{figure*}
\begin{figure*}[t!]
    \centering
    \includegraphics[width=0.9\textwidth]{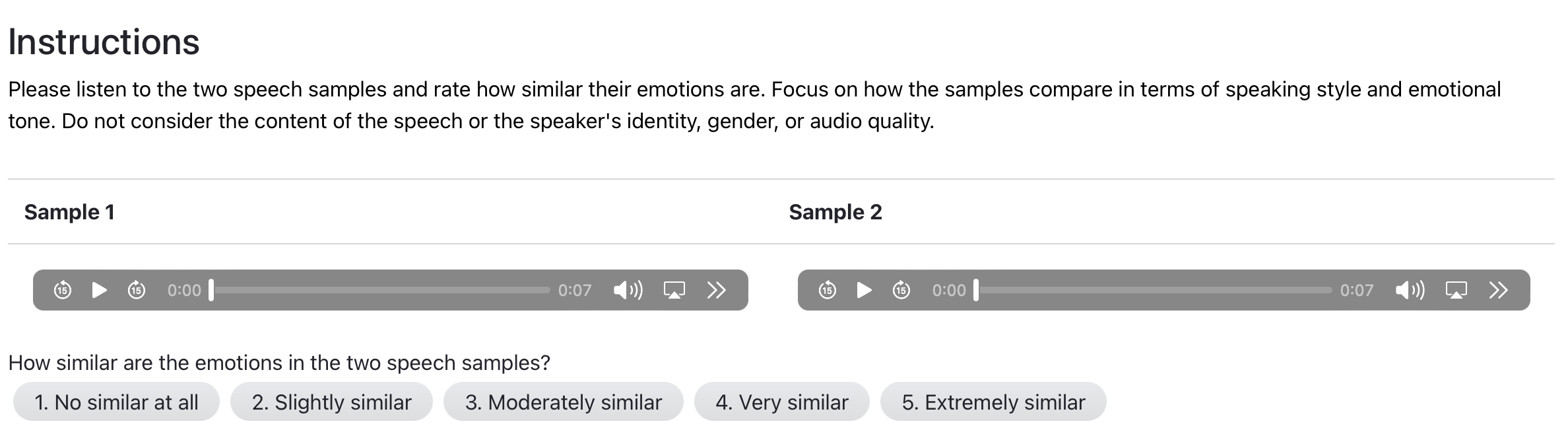}
    \caption{\textbf{Human evaluation.} Task description for the crowd-sourced raters to evaluate how close emotions in generated speech follows the ground truth.}
    \label{fig:mos_eval_setup_sim}
    \vspace{-0.4cm}
\end{figure*}

\textbf{Mean Opinion Score (MOS)}\label{app:subjective_evals}
We use crowd-sourcing to collect subjective ratings to evaluate the intelligibility, naturalness and synchronization of the generated speech. We use the same (randomly sampled) 50 videos from the test set of VoxCeleb2 (or LRS3)\footnote{Speakers in the test sets do not overlap with the speakers from the training sets.} for each model to generate speech. We then collect around seven ratings per video for each model. Overall, for both VoxCeleb2 and LRS3, we collect 4208 ratings from 387 different raters. The raters were English-speaking and were paid at least the minimum wage. They were crowd workers rather than domain experts in speech or audio. Ratings are absolute scores rather than A/B comparisons; the one comparative task is the emotion-similarity rating, in which two samples are played side by side (Figure ~\ref{fig:mos_eval_setup_sim}).

We present the raters with a generated speech (with volume normalization) overlayed with the original video or original video with original (or reconstructed) speech. 
We instruct raters to rate how natural speech in the video sounds, how intelligible (e.g. easy to understand) speech is in the video, and how synchronized the speech is with the video on a 1–5 scale, where 1 corresponds to very unnatural and 5 corresponds to very natural. In Figure~\ref{fig:mos_eval_setup} we show a screenshot seen by raters. The interfaces do not display system identity, so raters do not know which model produced a sample. Raters were recruited from the platform's English-language worker pool without additional domain screening. (Each rater was shown samples from all systems, with the presentation order randomized per rater.) Finally, we compute the MOS with confidence intervals calculated using bootstrap resampling with 10k iterations, providing a reliable estimate of the variability MOS results.

We further instruct raters to evaluate emotional consistency between video and generated speech ('video-speech emotions') and emotional expressiveness in speech ('speech emotions') by comparing ground truth and generated audios, see instructions in Figures~\ref{fig:mos_eval_setup_emotion} and~\ref{fig:mos_eval_setup_sim}. MOS results, Table~\ref{tab:voxceleb_mos_emotion},
highlight the advantages of visual conditioning for intelligibility, naturalness, synchronization, and emotional expressiveness. We use \textit{emotions2vec}\footnote{\url{https://huggingface.co/emotion2vec}} to select audio with non-neutral emotions (6 classes, each prob. $>99\%$).

\begin{table}[t!]
\caption{\textbf{Human evaluation on VoxCeleb2.} Mean Opinion Scores (MOS) (1-5) with 95\% confidence intervals for Intelligibility, Naturalness, and Synchronization. VTTS (VT-Scaled) achieves the best performance in Intelligibility (3.48) and Naturalness (3.20), while VTTS (TV-CoTemporal) performs best in Synchronization (2.50). Both outperform the TTS baseline across all metrics. Ground truth (GT) serves as a reference, and GT (discrete) is an upper bound due to speech discretization.}
\label{tab:voxceleb2_mos}
\vspace{0.2cm}
\centering
\resizebox{0.8\columnwidth}{!}{
\begin{tabular}{@{}lcccccc@{}}\toprule
    \textbf{Method} & \textbf{Intelligibility} ($\uparrow$) & \textbf{Naturalness} ($\uparrow$) & \textbf{Synchronization} ($\uparrow$) \\\midrule
    GT & 4.55 \textpm 0.09 & 4.79 \textpm 0.05 & 4.57 \textpm 0.10 &  \\
    GT (discrete) & 3.95 \textpm 0.13 & 3.77 \textpm 0.15 & 4.36 \textpm 0.12 &  \\ \midrule
    TTS (dmel ~\cite{bai2024dmel}) & 3.17 \textpm 0.19 & 2.92 \textpm 0.21 & 1.98 \textpm 0.15 & \\ \midrule
    \multicolumn{4}{@{}l}{\textbf{\textit{VTTS}}} \\
    Video-Causal-Streaming & 3.19 \textpm 0.17 & 2.99 \textpm 0.16 & 2.28 \textpm 0.17 &  \\
    TV-CoTemporal & \underline{3.35 \textpm 0.17}  & \underline{3.02 \textpm 0.19} & \textbf{2.50 \textpm 0.21} & \\
    VT-Scaled & \textbf{3.48 \textpm 0.15} & \textbf{3.20 \textpm 0.19} &
    \underline{2.48 \textpm 0.19} &  \\
\bottomrule
\end{tabular}
}
\vspace{-0.2cm}
\end{table}

\paragraph{MCD-DTW-SL vs. \alignmetric}~~~
MCD-DTW-SL measures spectral similarity by computing mel-cepstral distortion (MCD) after dynamic time warping between generated and reference speech. While effective for assessing acoustic fidelity, it operates at the frame level and allows non-linear temporal warping, which can mask synchronization errors by artificially aligning mis-timed speech. In contrast, \alignmetric operates at the phoneme level and explicitly measures absolute temporal deviation between corresponding phoneme centers without allowing time warping. As a result, \alignmetric directly captures perceptually salient audio-visual misalignment that DTW-based metrics may obscure, making it better suited for evaluating synchronization in video-conditioned speech generation.

\section{Implementation Details}\label{app:training_details}

{\bf Datasets}~~~
1) \textbf{LRS3}~\cite{afouras2018lrs3} is audio-visual dataset in English, compiled from TED and TEDx video presentations.
This dataset stands out for its focus on unconstrained long sentences, featuring a rich vocabulary of over 50k words and thousands of unique speakers. It contains approximately 151k videos with around 439h of speech with transcription.
There are 1,452 videos in the test split. 
2) \textbf{VoxCeleb2}~\cite{chung18b_interspeech} is a large-scale audio-visual dataset primarily designed for speaker recognition task but applicable to various audio-visual processing domains.
It consists of over 1M face-cropped YouTube videos from more than 6k distinct identities, resulting in 1.6k hours of speech \textit{w/o paired transcription}. 
The dataset is characterized by high variability in lighting conditions, image quality, pose, and motion blur, with an average video duration of 8s. This diversity in real-world conditions makes VoxCeleb2 particularly useful for developing robust models capable of performing well in unconstrained environments. Original data has video at 25fps (40ms per frame), or 25Hz, which we use for video representation extraction, while the audio is given at 16kHz and we extract speech representations at 40Hz (25ms per frame). speakers in train and test data do not intersect.

\paragraph{Hyper-Parameters Tuning} To select the best hyper-parameters we randomly sampled 2k samples from the training data and use them as the validation data throughout the training. After we find best hyper-parameters on the validation data, we retrain the final models including validation data into training data.

\vspace{-0.2cm}
\paragraph{Model Training} For our VTTS models we stack together speaker embedding, video, text and speech representations. Every modality has prepended begin of sentence representation ($<bos>$) and appended end of sentence representation ($<eos>$). 
Each modality's discrete values are mapped to a common dimension~$D'$ through their respective embedding layers and, optionally, additional linear projections before being fed to the decoder.
All our models have $\sim$250M parameters, with $D'=768$ and 36 transformer layers, following the Base architecture from~\cite{bai2024dmel}.
We follow masking strategy reported in~\cite{bai2024dmel}: for every training step with probability $p$ the sample in the minibatch is masked with the mean span of 3 tokens with masking ratio of 0.5.

We train final models using the AdamW optimizer with a learning rate of $4e-4$, learning rate warmup of $5$k steps, cosine learning rate schedule and gradient clipping of $1.0$.
We use dynamic batching to optimize the data packing with total batch size of 16.66 minutes.
We train all models till full convergence, with 3M maximum number of steps and with mixed precision training (BF16) on H100 GPUs with 80GB.
All models are trained with 8GPUs for 3-5 days.

\section{Additional LRS3 Evaluation Setting}
\label{app:lrs3_3to45}

The main paper reports primary cross-dataset comparisons on the full LRS3 test
set (Table~\ref{tab:lrs3_tab_obj}) for fairness. In addition, we introduce a
transfer-focused setting that keeps clips with durations between 3 and 45
seconds. We choose this range for two practical reasons: (i) clips shorter
than 3s tend to have high ASR WER due to ASR sensitivity to shorter audio as the lexical context is limited and ASR (whisper) is trained on 30s audio,
and (ii) clips
longer than 45s accumulate more ASR/forced-alignment drift, which can dominate
synchronization metrics. 
The same duration filter is applied uniformly to all
our models in Table~\ref{tab:lrs3_tab_obj_3to45}. This filter therefore also bounds \alignmetric's sensitivity to ASR and forced-alignment error, as discussed in Section ~\ref{sec:timesync_anl}.

\begin{table*}[t!]
\centering
\caption{\textbf{LRS3 transfer-focused setting.} This setting is
introduced in this work for detailed zero-shot transfer analysis. Unlike the
main full-set WER results, we additionally report \alignmetric in this setting.}
\label{tab:lrs3_tab_obj_3to45}
\resizebox{0.75\columnwidth}{!}{
\begin{tabular}{lccc}
\toprule
\textbf{Method} & \textbf{Training Dataset} & \textbf{WER (\%)} $\downarrow$ & \alignmetric (s) $\downarrow$ \\
\midrule
V2SFlow \citep{choi2025v2sflow} & LRS3 & 28.5 & - \\
LipVoicer \citep{yemini2024lipvoicer} & LRS3 & 21.4 & - \\
Lip2Speech \citep{kim2023lip} & LRS3 & 57.4 & - \\
SVTS \citep{mira2022svts} & LRS3 & 82.4 & - \\
VCA-GAN \citep{kim2021lip} & LRS3 & 90.6 & - \\
DiffV2S \citep{choi2023diffv2s} & LRS3 & 39.2 & - \\
VoiceCraft-Dub* (pretrained) \citep{sung2025voicecraft} & LRS3 fine-tuned & 1.38 & - \\
\midrule
TTS & VoxCeleb2 & \underline{5.3} & 0.34$\pm$0.28 \\
\multicolumn{4}{@{}l}{\textbf{\textit{VTTS}}} \\
TV-Global & VoxCeleb2 & 9.5 & 0.24$\pm$0.21 \\
VT-Scaled & VoxCeleb2 & 8.2 & 0.32$\pm$0.32 \\
VT-Global & VoxCeleb2 & 6.5 & 0.25$\pm$0.21 \\
TV-CoTemporal & VoxCeleb2 & \textbf{4.8} & \textbf{0.22$\pm$0.22} \\
\midrule
VT-Scaled & LRS3 & 6.3 & \underline{0.23$\pm$0.24}\\
TV-CoTemporal & LRS3 & 5.6 & \underline{0.23$\pm$0.22} \\
\bottomrule
\end{tabular}
}
\end{table*}

\section{External-Protocol VoxCeleb2 WER Reference}
\label{app:vdtts_external}

For completeness, we report published VoxCeleb2 WER numbers from
VDTTS~\citep{hassid2022vdtts} as reference-only values. These are not directly
comparable to our Table~\ref{tab:voxceleb2_oldnewdata} because the evaluation
protocol differs in data preprocessing/filtering, ASR backend, and transcript
pipeline.

\begin{table}[t]
\centering
\caption{\textbf{Reference-only VoxCeleb2-trained WER values.}
All rows are trained on VoxCeleb2 and evaluated on VoxCeleb2. Rows
from~\citep{hassid2022vdtts} use an external evaluation protocol; our best VoxCeleb2-trained non-global prefix variant is included for context from
Table~\ref{tab:voxceleb2_oldnewdata}. These rows are not directly comparable.}
\label{tab:vdtts_external_vox}
\resizebox{0.5\columnwidth}{!}{
\begin{tabular}{lc}
\toprule
\textbf{Method} & \textbf{WER (\%) $\downarrow$} \\
\midrule
VDTTS-VOXCELEB2~\citep{hassid2022vdtts} & 48.0 \\
VDTTS-LSVSR~\citep{hassid2022vdtts} & 25.0 \\
\midrule
VTTS (VT-Scaled, VoxCeleb2-trained, ours) & 12.2 \\
\bottomrule
\end{tabular}
}
\end{table}

\section{Video-to-Speech}
As one of the baselines we trained speech generation model conditioning only on the video input (no text input). The WER for this model is around 100\%, and MOS is 1.39~\textpm~0.10 for intelligibility, 1.60~\textpm~0.13 for naturalness and 1.49~\textpm~0.09 for synchronization. The interesting findings about this model are: a) the model is able to generate word $n$grams; b) the model is able to model properly the pauses and reflect the timing when people speaking or being silent.


\section{Qualitative Results}

\begin{figure*}[t]
    \begin{minipage}[t]{0.47\textwidth}
    \centering
    \includegraphics[width=0.9\columnwidth]{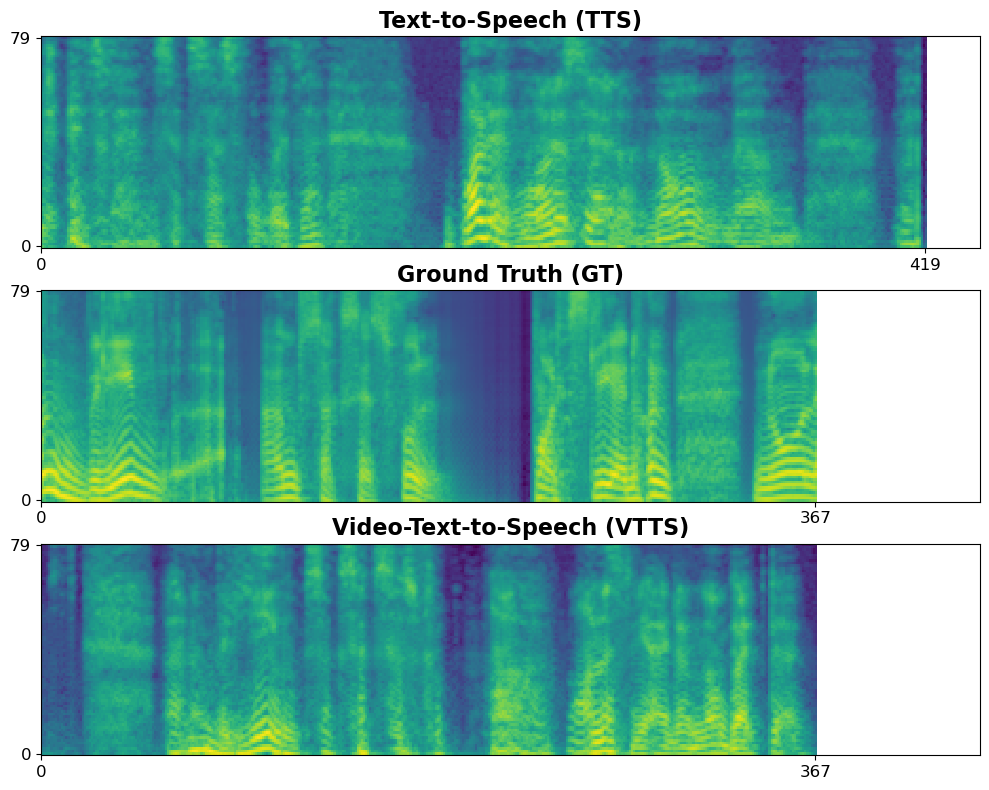}
    \caption{\textbf{Qualitative comparison of log mel-spectrograms.} Visualization of generated log mel-spectrograms: Text-to-Speech (TTS, top), Ground Truth (GT, middle), and our Video-Text-to-Speech (VTTS, bottom). VTTS (VT-Scaled) demonstrates better temporal alignment with GT (367 frames) compared to TTS (419 frames), showing the benefit of video conditioning for maintaining correct speech duration. The spectral patterns in VTTS (VT-Scaled) also more closely match GT's energy distribution.}
    \label{fig:qual_mel_suppl_v2}
    \end{minipage}
    \hfill
    \begin{minipage}[t]{0.47\textwidth}
    \centering
    \includegraphics[width=0.8\columnwidth]{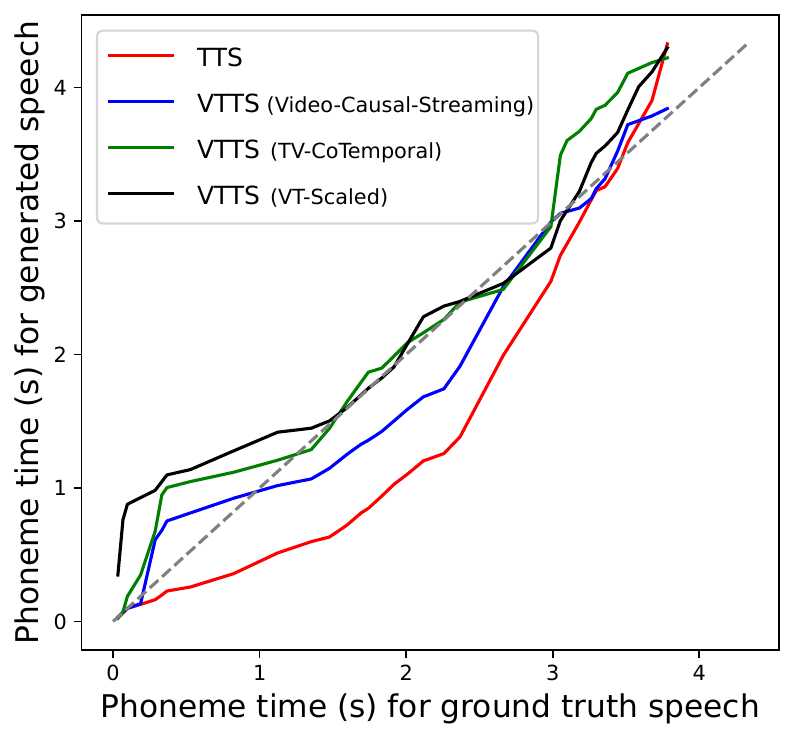}
    \caption{\textbf{Alignment between phonemes.} Temporal alignment visualization for example from Figure~\ref{fig:qual_mel_suppl_v2}. The plot compares phoneme timings between ground truth ($x$-axis) and generated speech ($y$-axis). Dashed gray line is the ideal time synchronization between GT and generated speech. TTS is way out for the proper timing compared to VTTS (TV-CoTemporal) and VTTS (VT-Scaled).}
    \label{fig:metric_example_suppl}
    \end{minipage}
\end{figure*}

\begin{figure*}[t]
\begin{minipage}[t]{0.47\textwidth}
    \centering
    \includegraphics[width=0.9\columnwidth]{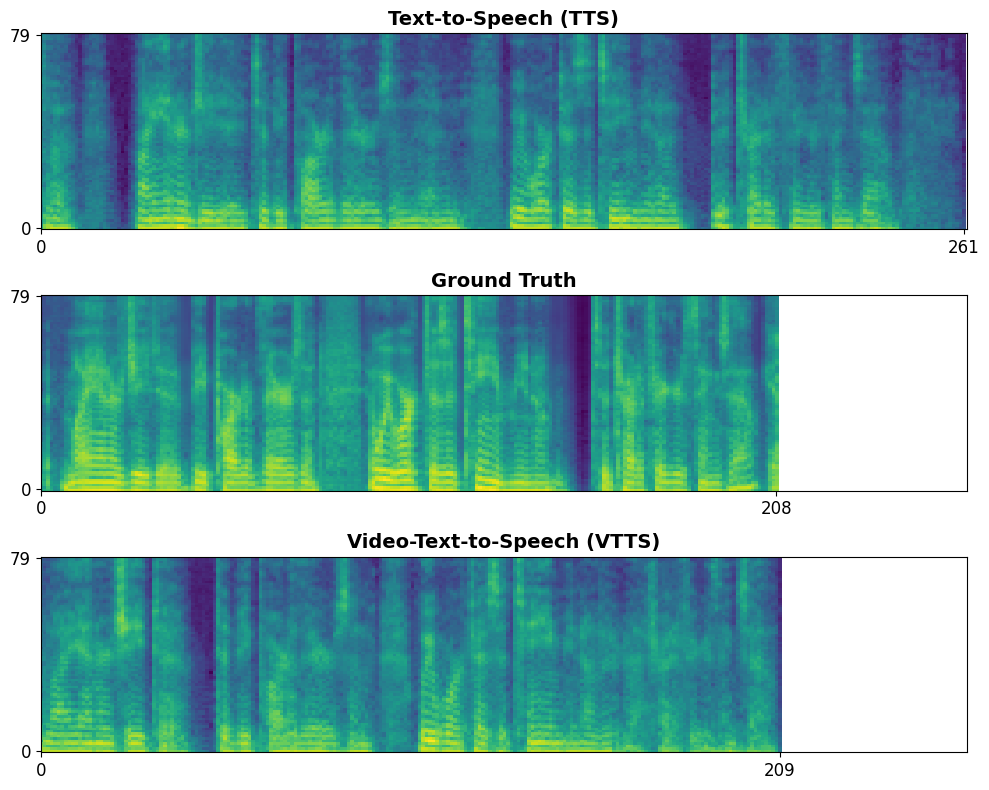}
    \caption{\textbf{Qualitative comparison of log mel-spectrograms.} Visualization of generated log mel-spectrograms from different methods: Text-to-Speech (TTS, top), Ground Truth (GT, middle), and our Video-Text-to-Speech (VTTS, bottom). VTTS (VT-Scaled) demonstrates better temporal alignment with GT (208 frames) compared to TTS (261 frames), showing the benefit of video conditioning for maintaining correct speech duration. The spectral patterns in VTTS (VT-Scaled) closely match GT's harmonic structure and energy distribution, particularly visible in the lower frequency bands (yellow regions). Additionally, VTTS (VT-Scaled) accurately captures the temporal dynamics of speech, including pauses and intensity variations, leading to more natural speech generation.}
    \label{fig:qual_mel_suppl_164}
\end{minipage}
\hfill
\begin{minipage}[t]{0.47\textwidth}
    \centering
    \includegraphics[width=0.8\columnwidth]{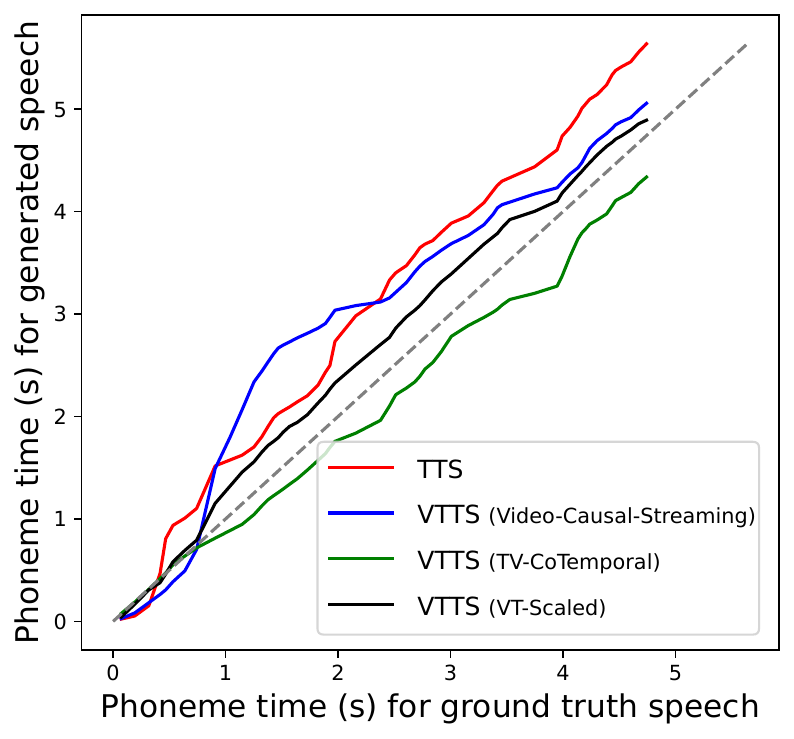}
    \caption{\textbf{Alignment between phonemes.} Temporal alignment visualization for example from Figure~\ref{fig:qual_mel_suppl_164}. The plot compares phoneme timings between ground truth ($x$-axis) and generated speech ($y$-axis). VTTS demonstrate superior temporal alignment by following the ideal synchronization line (dashed diagonal) more closely than TTS, which shows significant temporal drift. This example highlights how video conditioning helps maintain proper speech timing.}
    \label{fig:metric_example_suppl_164}
\end{minipage}
\end{figure*}

\begin{figure*}[h!]
\begin{minipage}[t]{0.47\textwidth}
    \centering
    \includegraphics[width=0.9\columnwidth]{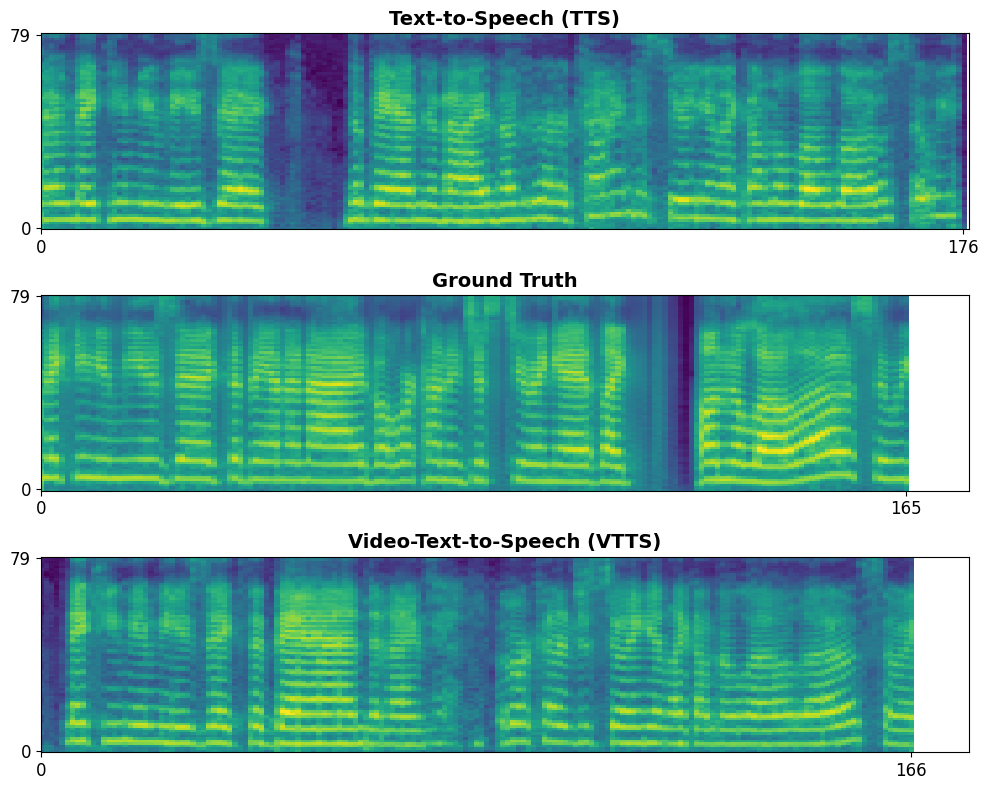}
    \caption{\textbf{Failure case analysis of log mel-spectrograms.} Visualization of generated log mel-spectrograms from different methods: Text-to-Speech (TTS, top), Ground Truth (GT, middle), and our Video-Text-to-Speech (VTTS, bottom). While VTTS (VT-Scaled) maintains better temporal alignment with GT (165 frames vs TTS's 176 frames), both VTTS (VT-Scaled) and TTS struggle to accurately capture GT's harmonic structure and energy distribution. Despite having video conditioning, VTTS (VT-Scaled) shows degraded spectral quality particularly in the mid-frequency ranges, though it still preserves some temporal speech dynamics like pauses. This example highlights current limitations in generating complex spectral patterns.}
    \label{fig:qual_mel_suppl_490}
\end{minipage}
\hfill
\begin{minipage}[t]{0.47\textwidth}
    \centering
    \includegraphics[width=0.8\columnwidth]{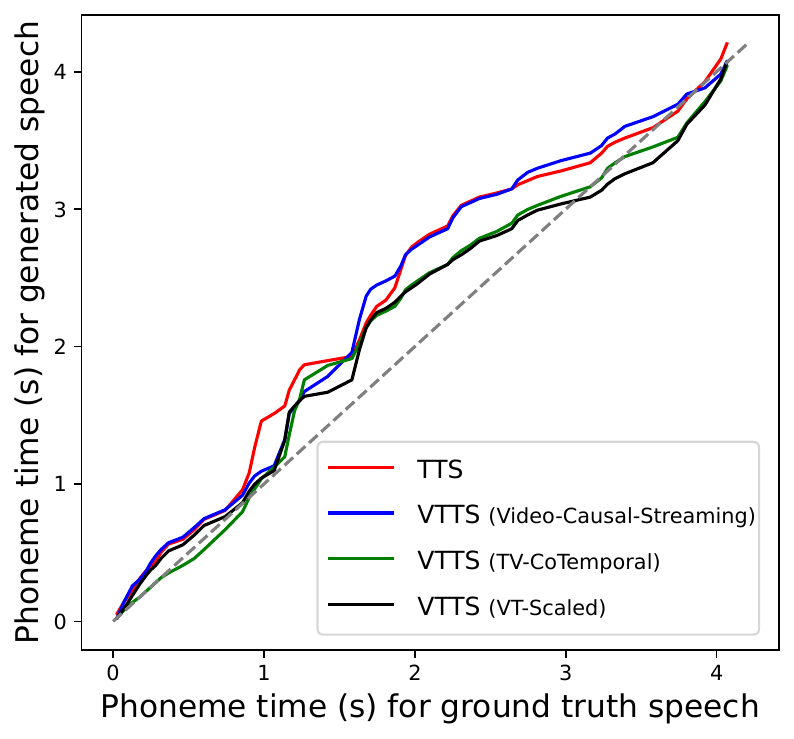}
    \caption{\textbf{Alignment between phonemes.} Temporal alignment visualization for failure case corresponding to Figure~\ref{fig:qual_mel_suppl_490}. Plot shows phoneme timing comparison between ground truth ($x$-axis) and generated speech ($y$-axis). While VTTS maintain better alignment than TTS, all models show deviation from ideal synchronization (dashed diagonal), particularly in later segments, illustrating challenges in maintaining temporal coherence for complex speech patterns.}
    \label{fig:metric_example_suppl_490}
\end{minipage}
\end{figure*}

In Figures~\ref{fig:qual_mel_suppl_v2},~\ref{fig:qual_mel_suppl_164}, and~\ref{fig:qual_mel_suppl_490}, we show log mel-spectrogram comparisons between TTS, Ground Truth (GT), and our VTTS (VT-Scaled) model across different scenarios. These visualizations include both successful cases where VTTS  (VT-Scaled) effectively captures temporal dynamics and spectral patterns, and failure cases (Figure ~\ref{fig:qual_mel_suppl_490}) that highlight current limitations. Through these examples, we can analyze how video conditioning helps maintain proper speech duration and temporal alignment, while also identifying challenges in generating complex spectral information. Furthermore, to analyze temporal synchronization between generated and ground truth speech, we visualize phoneme-level alignments in Figures~\ref{fig:metric_example_suppl}, \ref{fig:metric_example_suppl_164}, and \ref{fig:metric_example_suppl_490}. Each plot shows the relationship between phoneme timings in ground truth ($x$-axis) versus generated speech ($y$-axis), where perfect synchronization would follow the diagonal dashed line. The different variants of our model, VTTS (VT-Scaled) consistently demonstrate better temporal alignment compared to TTS, as evidenced by their closer distance to the ideal diagonal. This visualization helps quantify how video conditioning helps to maintain proper speech timing and rhythm, with VTTS (VT-Scaled) variants showing improved temporal coherence across different examples. 

\clearpage

\end{document}